%% file: tokenmixer-large.tex
\documentclass[sigconf]{acmart}

\AtBeginDocument{%
  }

\usepackage{booktabs}
\usepackage{multirow}
\usepackage{algorithm}
\usepackage{algorithmic}
\usepackage{enumitem}
\usepackage{xcolor}
\usepackage{pifont}
\usepackage{subfigure}
\usepackage{tabularx}
\usepackage{float}
\usepackage{threeparttable} 

\usepackage{makecell}
\usepackage{siunitx}    
\usepackage{graphicx}   
\usepackage{subcaption} 
\usepackage{listings}
\usepackage{mathtools}

\definecolor{codegray}{rgb}{0.95,0.95,0.95}
\definecolor{commentteal}{rgb}{0.0, 0.5, 0.5} 
\definecolor{keywordpurple}{rgb}{0.58,0,0.82}

\lstdefinestyle{tfstyle}{
    backgroundcolor=\color{codegray},
    commentstyle=\color{commentteal}\itshape,
    keywordstyle=\color{keywordpurple}\bfseries,
    basicstyle=\ttfamily\footnotesize,
    breakatwhitespace=false,
    breaklines=true,
    captionpos=t,
    keepspaces=true,
    frame=lines,
    language=Python,
    morekeywords={all2all, pertokenOperation, transpose},
    columns=flexible,
}

\begin{document}

\title[TokenMixer-Large]{TokenMixer-Large:  Scaling Up Large Ranking Models in Industrial Recommenders}

\author{Yuchen Jiang$^{*1}$, Jie Zhu$^{*1}$, Xintian Han$^{*1}$, Hui Lu$^{*1}$, Kunmin Bai$^{*1}$, Mingyu Yang$^{*2}$, Shikang Wu$^{*2}$\\Ruihao Zhang$^{*1}$, Wenlin Zhao$^{*2}$, Shipeng Bai$^{1}$, Sijin Zhou$^{2}$, Huizhi Yang$^{2}$, Tianyi Liu$^{1}$, Wenda Liu$^{1}$\\ Ziyan Gong$^{1}$, Haoran Ding$^{1}$,  Zheng Chai$^{1}$, Deping Xie$^{1}$, Zhe Chen$^{\dagger1}$, Yuchao Zheng$^{\dagger1}$, Peng Xu$^{1}$}
\affiliation{
	\country{ByteDance AML$^{1}$, ByteDance$^{2}$\\
    \{jiangyuchen.jyc, zhujie.zj, hanxintian, luhui.xx, baikunmin.by, yangmingyu, wushikang, zhangruihao.711, zhaowenlin, baishipeng, zhousijin, yanghuizhi, liutianyi.2024, liuwenda, gongziyan, dinghaoran, chaizheng.cz, deping.xie, chenzhe.john,  zhengyuchao.yc, xupeng\}@bytedance.com}}

\thanks{* Equal Contributions. \\ $\dagger$ Corresponding authors.}

\renewcommand{\shortauthors}{Yuchen Jiang et al.}

\begin{abstract}

While scaling laws for recommendation models have gained significant traction, existing architectures such as Wukong, HiFormer, and DHEN often struggle with sub-optimal designs and hardware under-utilization, limiting their practical scalability. Our previous TokenMixer architecture (introduced in the RankMixer paper) balanced effectiveness and efficiency by replacing self-attention with a lightweight token mixing operator; however, it encountered critical bottlenecks in deeper configurations, including sub-optimal residual paths, vanishing gradients, incomplete MoE sparsification, and constrained scalability. In this paper, we propose \textbf{TokenMixer-Large}, a systematically evolved architecture designed for extreme-scale recommendation. By introducing a mixing-and-reverting operation, inter-layer residuals, and an auxiliary loss, we ensure stable gradient propagation as model depth increases. Furthermore, we incorporate a Sparse Per-token MoE to enable efficient parameter expansion. TokenMixer-Large successfully scales to \textbf{7 billion} parameters for online traffic and \textbf{15 billion} in offline experiments. Currently deployed in multiple scenarios at ByteDance, TokenMixer-Large has achieved substantial performance gains: delivering a 1.66\% increase in orders and a 2.98\% boost in per-capita preview payment GMV for e-commerce; improving ADSS by 2.0\% in advertising; and achieving 1.4\% revenue growth in live streaming.
\end{abstract}






\maketitle

\input{text/introduction}

\input{text/related_work}

\input{text/method}

\input{text/experiment}

\input{text/conclusion}

\bibliographystyle{ACM-Reference-Format}
\balance
\bibliography{software}

\input{text/appendix}

\end{document}

%% file: text/introduction.tex
\vspace{-0.2cm}
\section{Introduction}
Recommendation systems~(RS) play a pivotal role in today's internet ecosystem, powering critical applications such as e-commerce~\cite{chen2019behavior,din}, online advertising~\cite{lin2023can,jiang2022adaptive}, content feeds ~\cite{zhu2025rankmixer,chang2023twin}, and live streaming\cite{longer}. Consequently, both industry and academia are engaged in continuous iteration and upgrades of recommendation models and architectures. With the widespread adoption of deep learning, Deep Learning Recommendation Models (DLRMs) have demonstrated remarkable effectiveness. These models typically first transform high-dimensional sparse user, item, and behavioral features into low-dimensional dense embeddings via an embedding layer. Subsequently, sophisticated feature interaction layers are designed to capture potential user-item relationships.

Inspired by the success of Large Language Models (LLMs)~\cite{achiam2023gpt,touvron2023llama,liu2024deepseek}, research in feature interaction has begun exploring the scaling laws of DLRMs: investigating whether continuous performance gains can be achieved by increasing model parameters and FLOPs. Early efforts focused primarily on merely expanding model width or interaction layer parameters, lacking thoughtful architectural design, which limited their effectiveness. Subsequent works like Wukong~\cite{zhangwukong}, HiFormer~\cite{gui2023hiformer}, and DHEN~\cite{zhang2022dhen} attempted to design more refined model structures and scale them (in size, width and depth), but often overlooked hardware-aligned co-design and optimization, leading to suboptimal performance. 

Currently, the leading SOTA model structure in industry is TokenMixer—a highly simplified variant of the Transformer. It significantly reduces computational complexity by replacing attention with a lightweight token mixing operation. RankMixer~\cite{zhu2025rankmixer} employed it as a backbone network and verified the effectiveness of the structure in the ranking model. Through hardware-aware co-design, it substantially improves Model FLOPs Utilization (MFU) while maintaining scaling efficiency, greatly enhancing hardware usage.
Despite its industrial success (e.g., RankMixer), we identify several key limitations in TokenMixer:
\begin{itemize}[topsep=0.5pt, leftmargin=*]
    \item \textbf{Sub-optimal Residual Design}: Since RankMixer transforms original tokens into new tokens through a mixing mechanism, it requires the number of new tokens $T'$ matches the original number $T$ to ensure the integrity and propagation of the original semantic information. Furthermore, in its add \& norm operations, RankMixer directly adds tokens before and after mixing, which often triggers semantic misalignment, ultimately leading to sub‑optimal performance.
    \item \textbf{Impure Model Architecture}: Due to historical iterative processes, many deployed models retain numerous fragmented operators (e.g., LHUC~\cite{swietojanski2016learning}, DCNv2~\cite{wang2021dcn}) even after adopting TokenMixer as the backbone. These memory-bound operators, characterized by low computational intensity but high memory access and communication overhead, degrading the overall model MFU.
    \item \textbf{Insufficient Gradient Updates in Deep Models}: TokenMixer is typically configured with a shallow depth (e.g., 2 layers in RankMixer) in industrial settings. As model depth increases, achieving training stability and performance gains becomes challenging. The current TokenMixer lacks designs tailored for deep architectures.
    \item \textbf{Inadequate MoE Sparsification}: While RankMixer employs Sparse MoE~\cite{fedus2022switch} to optimize model cost, it relies on a "Dense Train, Sparse Infer" paradigm, failing to reduce training costs. Moreover, the ReLU-MoE~\cite{Wang2024remoe} used exhibits activation dynamism, making it difficult to predict the number of experts activated per batch during inference. This necessitates truncation or fallback strategies for activated experts, which is not inference-friendly.
    \item \textbf{Limited Scaling Exploration}: Due to the framework and training efficiency constraints, RankMixer’s parameter scale was only pushed to about 1-Billion parameters.
\end{itemize}
To address these challenges, we design and thoroughly explore TokenMixer-Large. The TokenMixer-Large framework comprises modules like Tokenization, TokenMixer-Large Block, and Sparse-Pertoken MoE. A TokenMixer-Large Block contains components such as Mixing \& Reverting, Pertoken-SwiGLU, Residual \& Normalization, and Interval Residual \& Residual Loss. By stacking multiple TokenMixer-Large Blocks, we significantly improve offline and online model performance in industrial scenarios.
The contributions of this paper are summarized as follows:
\begin{itemize}[topsep=0.5pt, leftmargin=*]
    \item We propose TokenMixer-Large. By revisiting and redesigning the flawed residual mechanism in TokenMixer, we introduce the Mixing \& Reverting operation. Extensive ablation studies on key components demonstrate that TokenMixer-Large is a superior model architecture.
    \item We observe that as models scale up, the benefits provided by low-level fragmented operators can be subsumed by stacking multiple TokenMixer Blocks. Removing these operators drastically improves model MFU and hardware utilization. We integrate this "pure model" philosophy into TokenMixer-Large, elevating the MFU of the advertising backbone network to 60\%
    \item To tackle the issue of insufficient gradient updates in deep models, we incorporate interval residuals, an auxiliary loss mechanism and down-matrix small initialization within the TokenMixer-Large Block, facilitating better convergence for deep models.
    \item We upgrade the relu-MoE in RankMixer(TokenMixer) to a Sparse-Pertoken MoE version with a unified "Sparse Train, Sparse Infer" paradigm. Coupled with efficient operator design and engineering optimizations such as FP8 and \textbf{Token Parallel}, Sparse-Pertoken MoE significantly reduces both training and inference costs.
    \item We conduct a comprehensive study on the scaling laws of different models, validating the superiority of TokenMixer-Large. We successfully scaled the model to \textbf{15B} and \textbf{7B} parameters in the offline experiments for the Douyin ads and e-commerce scenarios, respectively. In the online traffic experiments, these were scaled to \textbf{7B} and \textbf{4B} parameters, respectively. TokenMixer-Large is now deployed across multiple online scenarios at ByteDance, achieving online gains of 2\% ADSS and 2.98\% GMV while serving hundreds of millions of users.
\end{itemize}

%% file: text/related_work.tex
\vspace{-0.2cm}
\section{Related Work}
\label{sec:related work}

The core of deep learning recommendation models (DLRMs)~\cite{covington2016deep} lies in designing effective feature interaction architectures. Wide and Deep~\cite{cheng2016wide}, as one of the classical works, captures both low‑order and high‑order feature interactions through a bagging‑style approach. This modeling paradigm is reflected in subsequent designs such as DeepFM~\cite{guo2017deepfm}, XDeepFM~\cite{lian2018xdeepfm}, PNN~\cite{qu2018product}, DCN~\cite{wang2017deep}, DCNv2~\cite{wang2021dcn}, and FCN~\cite{cheng2016wide}. However, being early‑stage designs, these models still rely on inefficient, high‑latency operators from the CPU era and fail to fully leverage the computational power of GPUs. Building on the bagging of different substructures, DHEN~\cite{zhang2022dhen} and Wukong~\cite{zhangwukong} further validated the scaling laws in recommendation systems, demonstrating that a "unified design, stacked layers" iteration pattern can also achieve significant gains in recommendation tasks.


As a foundational model, the Transformer~\cite{transformer} has been widely proven in the GPU era to possess strong versatility (CV~\cite{dosovitskiy2020image, liu2021swin}, NLP~\cite{touvron2023llama, liu2024deepseek} and RS~\cite{autoint}) and favorable scaling‑law~\cite{kaplan2020scaling, wu2024performance} properties. In the recommendation domain, models such as AutoInt~\cite{autoint} and HiFormer~\cite{gui2023hiformer} enhances feature interaction capabilities by incorporating attention mechanisms. RankMixer~\cite{zhu2025rankmixer} and MLP-Mixer~\cite{tolstikhin2021mlp} further simplifies attention into a more lightweight token‑mixing operation, reducing the quadratic computational cost associated with standard attention. Works such as LONGER~\cite{longer}, HSTU~\cite{zhaiactions}, and MTGR~\cite{han2025mtgr} have also demonstrated, to varying degrees, the scaling potential of Transformer‑based models (and their variants) in sequential modeling and generative recommendation scenarios.

%% file: text/method.tex
\vspace{-0.2cm}
\section{Methodology}
\label{sec:methods}

\begin{figure*}[htbp]
    \centering
    \includegraphics[width=0.80\linewidth]{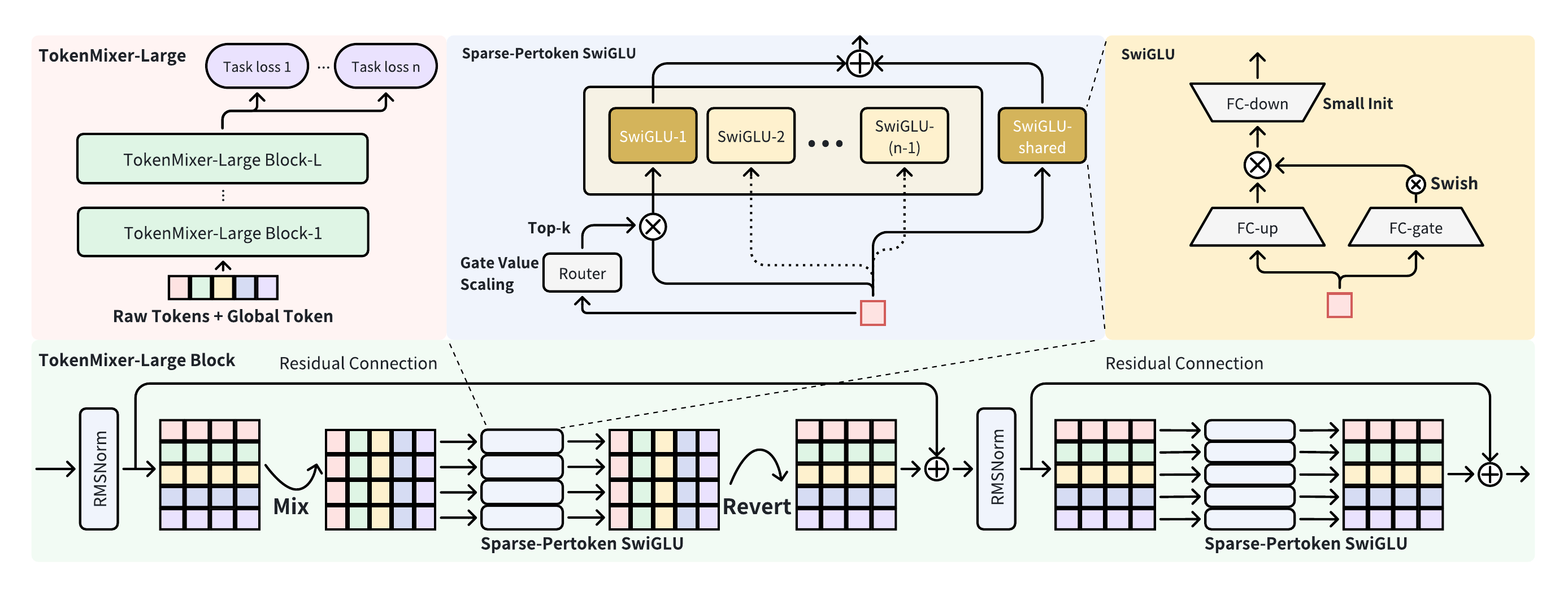}
    \captionsetup{skip=0pt}
    \caption{The architecture of the TokenMixer-Large. Raw tokens include all original features as well as features from sequence aggregation and extraction (such as din~\cite{din}/ longer~\cite{longer}). The entire Tokenmixer-Large model consists of multiple Tokenmixer-Large Blocks, and the backbone of each block consists of (Norm, Mixing, S-P~MoE, Reverting, Norm, S-P~MoE) and residuals.}
    \label{fig:arch_overall}
    \vspace{-0.2cm}
\end{figure*}


 \subsection{Overall Architecture}
As illustrated in Figure~\ref{fig:arch_overall}, The overall architecture of the TokenMixer-Large framework consists of three parts: 1).Tokenization, which transfers raw highly sparse one-hot features to dense embeddings, then to dimension-aligned semantic tokens. 2). Token Mixing \& Channel Mixing, which follows a "Mixing and Reverting" paradigm to solve the dimension mismatch problem in RankMixer(TokenMixer). 3). Sparse-Pertoken MoE, which is the upgraded version of Pertoken-FFN/relu-MoE mentioned in RankMixer(TokenMixer). Finally, we use a mean pooling method to aggregate output tokens, and the output representation will be used for various tasks prediction.

\subsection{Tokenization}
\subsubsection{Semantic Group-wise Tokenizer}
Industrial recommendation systems usually include user features, item features, sequence features (e.g., user behavior history covering short-term~(DIN~\cite{din}), long-term~(SIM~\cite{pi2020search}), ultra long-term~(LONGER~\cite{longer}) interests) and cross features. The first step involves projecting all these features from a high-dimensional sparse space into a low-dimensional dense space via an embedding layer, where Feature $F_i$ will be embedded to embedding $e_i$ with varying dimension $d_i$
\begin{equation} e_{i} = \text{Embedding}(F_i,~d_i)  \in \mathbb{R}^{d_i}\end{equation}
Given that TokenMixer-Large is a highly parallel architecture design, it is necessary to transform variable-length embedding representations into dimension-aligned feature tokens to facilitate subsequent computations. To achieve this, we group embeddings of different dimensions according to their semantic meanings and concatenate the embeddings within each group $\{G_1,~ ...,~G_{T-1}\}$ before performing compression and alignment. Considering that different semantic groups represent distinct feature characteristics, we employ different DNN mappings during the compression and alignment process to preserve their heterogeneity.
\begin{equation} \mathrm{X}_{i} = \text{MLP}_{i}(\text{concat}[e_{l},~...,~e_{m}]) , ~~~ e_{l},...,e_{m}  \in  G_{i}\end{equation}

\subsubsection{Global Token}
Beyond semantically grouped tokens, we also introduce a \textbf{global token} specifically designed to encapsulate global information. Drawing inspiration from the \texttt{[CLS]} token mechanism in BERT~\, this global token is designed to aggregate global information and propagate it to other tokens.
\begin{equation} \mathrm{X_{G}} = \text{MLP}_{g}(\text{concat}\bigl[G_{1},~...,~G_{T-1}\bigr]) \end{equation}
The final input of TokenMixer-Large is the combination of the global token and enhanced tokens. 


\begin{equation} 
\mathbf{X} = \text{concat}\bigl[\mathrm{X_{G}},\mathrm{{X}}_0,~...,~\mathrm{{X}}_{T-1}\bigr] \in \mathbb{R}^{T \times D}
\end{equation}
  
\subsection{TokenMixer-Large Block}
TokenMixer-Large employs a stacked structure of multiple TokenMixer-Large blocks. Each block can be decomposed into three components: mixing module, pertoken SwiGLU, and normalization.
  
\subsubsection{Mixing \& Reverting}
A RankMixer~(TokenMixer) block can be formulated as follows.

\begin{align}
    &[\dots,[\mathrm{x}_{t}^{(0)}, \dots, \mathrm{x}_{t}^{(H)}],\dots] = \text{split}(\mathbf{X}) \in \mathbb{R}^{T\times H\times(D/H)} \\
    &\mathrm{H}_{h} = \text{concat}[\mathrm{x}_{1}^{(h)}, \dots, \mathrm{x}_{T}^{(h)}] \in \mathbb{R}^{T\times D/H} \\
    &\mathbf{H} = \text{concat}[\mathrm{H}_{1}, \dots, \mathrm{H}_{h}] \in \mathbb{R}^{H\times (T \cdot D/H)} \\
    &\mathbf{H}^{\text{next}} = \text{Norm}(\text{pSwiGLU}(\mathbf{H}) + \mathbf{H}) \in \mathbb{R}^{H\times (T * D/H)}
\end{align}

As shown, the number of tokens in the original input is $T$, while the number of tokens after mixing in the i-th layer is H. Unless each layer's $H$ remains the same and $T$ is equal to $H$, residual connections cannot propagate smoothly across layers. To overcome this limitation, we design TokenMixer-Large, featuring a special two-layer TokenMixer structure: the \textbf{first layer} is responsible for mixing information among the original tokens, and \textbf{the second layer} specifically restores the dimension of the mixed tokens. This symmetric "mixing–reverting" design ensures dimensional consistency of the input and output across modules, thereby establishing a continuous signal pathway from the initial input to the deep network. It enables stable residual connections at every layer and effectively prevents the loss of gradient information. Detailed comparisons with RankMixer(TokenMixer) can be found in section~\ref{sec:Detailed_Comparison_With_Rankmixer}


\begin{align}
    &[...,[\mathrm{x}_{t}^{(1)}, \mathrm{x}_{t}^{(2)}, \dots, \mathrm{x}_{t}^{(H)}],...] = \text{split}(\mathrm{\mathbf{X}}) \in \mathbb{R}^{T\times H\times(D/H)} \\
    &\mathrm{H_{h}} = \text{concat}[\mathrm{x}_{1}^{(h)}, \mathrm{x}_{2}^{(h)}, \dots, \mathrm{x}_{T}^{(h)}] \in \mathbb{R}^{ (T*D/H)} \\
    &\mathrm{\mathbf{H}} = \text{concat}[\mathrm{H}_{1}, \mathrm{H}_{2}, \dots, \mathrm{H_{H}}]
    \in \mathbb{R}^{H\times (T*D/H)} \\
    &\mathrm{\mathbf{H}^{next}} = \text{Norm}(\mathrm{pSwiGLU}(\mathrm{\mathbf{H}}) + \mathrm{\mathbf{H}}) \in \mathbb{R}^{H\times (T*D/H)} \\
    &[...,[\mathrm{x'}_{1}^{(h)}, \mathrm{x'}_{2}^{(h)}, \dots, \mathrm{x'}_{T}^{(h)}],...] = \text{split}(\mathrm{\mathbf{H}^{next}}) \in \mathbb{R}^{T\times H\times(D/H)}\\
    &\mathrm{X_{t}^{revert}} = \text{concat}[\mathrm{x'}_{t}^{(1)}, \mathrm{x'}_{t}^{(2)}, \dots, \mathrm{x'}_{t}^{(H)}] \in \mathbb{R}^{D}\\
    &\mathrm{\mathbf{X}^{revert}} = \text{concat}[\mathrm{X_{1}^{revert}}, \mathrm{X_{2}^{revert}}, \dots, \mathrm{X_{T}^{revert}}] \in \mathbb{R}^{T\times D} \\    &\mathrm{\mathbf{X}^{next}} = \text{Norm}(\mathrm{pSwiGLU}(\mathrm{\mathbf{X}^{revert}}) + \mathrm{\mathbf{X}}) \in \mathbb{R}^{T\times D}
\end{align}

\subsubsection{Pertoken SwiGLU}
RankMixer(TokenMixer) introduces a parameter isolated feed‑forward network architecture termed pertoken FFN, designed to model the feature heterogeneity among different tokens. In TokenMixer-Large, we upgrade the pertoken FFN to a pertoken SwiGLU.
\begin{equation}
    \mathrm{pSwiGLU(\cdot)} = \mathrm{FC_{\text{down}} \left({\rm Swish} \left(FC_{\text{gate}} \left(\cdot \right) \right)\odot FC_{\text{up}} \left(\cdot \right)  \right)},
\end{equation}
where
\begin{equation}
\mathrm{FC_i}(\mathbf{x}) = \mathrm{W}_i^t \mathrm{x}_{t} + \mathrm{b}_i^t, \quad i \in \{\text{up},\text{gate},\text{down}\}
\end{equation}
is the single MLP of the pertoken SwiGLU, 
$\{\mathrm{W}_{\text{up}}^{t},\mathrm{W}_{\text{gate}}^{t}\} \in \mathbb{R}^{D \times nD}$, 
$\{\mathrm{b}_{\text{up}}^{t},\mathrm{b}_{\text{gate}}^{t}\} \in \mathbb{R}^{nD}$,
$\mathrm{W}_{\text{down}}^{t} \in \mathbb{R}^{nD \times D}$, 
$\mathrm{b}_{\text{down}}^{t} \in \mathbb{R}^{D}$,
$n$ is a hyperparameter to adjust the hidden dimension of the pertoken SwiGLU,
${\rm Swish(\cdot)}$ is the Swish activation function, $\mathbf{x}_t \in \mathbb{R}^{D}$ is the $t$-th token.



\subsubsection{Residual \& Normalization}
Standard residuals are shown in Figure~\ref{fig:arch_overall}. In addition, we replace the standard LayerNorm with the more lightweight RMSNorm~\cite{zhang2019root}. By removing the mean-centering step in LayerNorm, RMSNorm preserves core normalization capabilities while significantly reducing computational overhead. However, under the Post-Norm design in RankMixer(TokenMixer), we often encounter issues of numerical explosion and gradient instability. To address these stability concerns, we adopt a \textbf{Pre-Norm} design to replace the original \textbf{Post-Norm} architecture. Detailed comparasion can be found in Appendix~\ref{sec:Normalization_Position}

\subsubsection{Inter-Residual \& Auxiliary Loss}

\begin{figure}
    \centering
    \includegraphics[width=\linewidth]{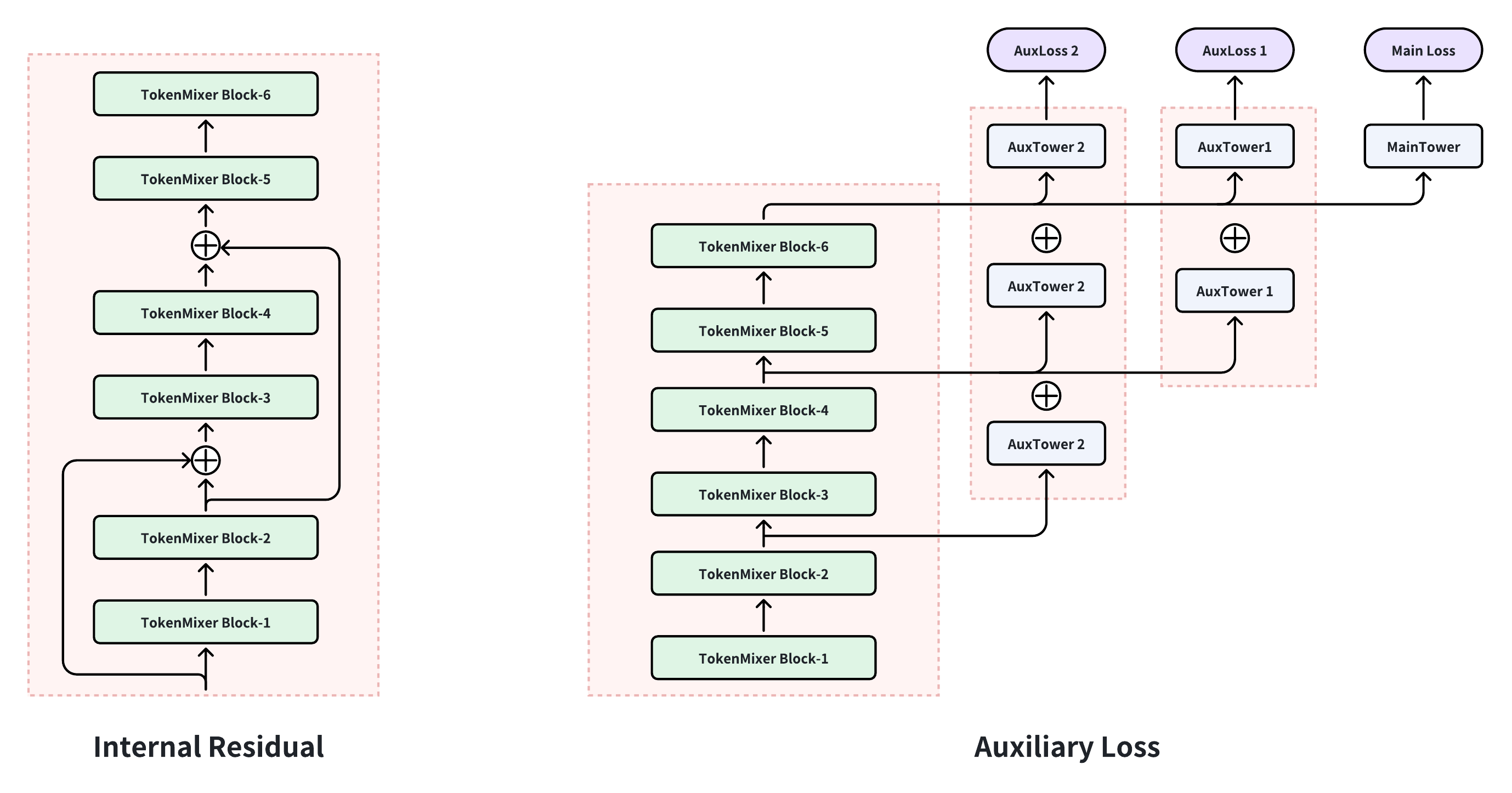}
    \vspace{-0.3cm}
    \captionsetup{skip=-2pt}
    \caption{Internal Residual and Auxiliary Loss}
    \label{fig:interRes}
\end{figure}

As Figure~\ref{fig:interRes} shown, in addition to the standard residual connections, we incorporate \textbf{inter-residual connections} (typically added at intervals of 2 or 3 layers) to strengthen information flow across layers. It primarily addresses 
the gradient vanishing problem. By enhancing the transmission of lower-layer features to higher layers, it accelerates the convergence of parameters in early layers and alleviates the gradual decay of gradients in deep networks.
Interval-residual connections are \textbf{not} recommended for the final layer. The primary role of the last layer is to distill highly abstract, high-level features to provide precise support for subsequent classification tasks. Introducing excessive raw lower-level information could interfere with this abstraction process and thereby hurt final task performance.

Meanwhile, by combining the logits computed from the outputs of the lower-layers with those of the higher layers for the calculation of joint loss, a lightweight \textbf{auxiliary loss} is formed. This mechanism essentially enables the lower layers to learn to "estimate deviations in higher-layer features," thereby strengthening their feature representation capacity. It helps prevent insufficient training of lower-layer parameters as network depth increases, leading to more thorough parameter learning and more stable performance across the deep model.

\vspace*{-5pt}
\vspace*{-5pt}

\subsection{Sparse-Pertoken MoE}

To further enhance the cost-effectiveness of TokenMixer-Large, we design a Sparse-pertoken MoE, which expands each expert into multiple sub-experts based on the pertoken SwiGLU and activates them sparsely. This enables "\emph{\textbf{sparse training and sparse serving}}" compared with Rankmixer's "dense training and sparse serving", significantly reducing the training and inference costs of very large-scale recommendation models.

\subsubsection{First Enlarge, Then Sparse} In terms of iteration strategy, we adopt a "first enlarge, then sparse" approach: first scaling up the model to gain effectiveness, then splitting the pertoken SwiGLUs and activating them sparsely to obtain efficiency benefits. The \textbf{Sparse-Pertoken MoE} (S-P MoE) can be formulated as follows:
\begin{equation}
\begin{aligned}
\mathrm{S\textbf{-}P~MoE(\cdot)} &=\sum_{j=1}^{k} g_{j}(\cdot)\cdot \text{Expert}_{j}(\cdot),~ \textbf{if j is chosen},\\
      \text{Expert}_{j}(\cdot)&= \mathrm{FC_{\text{down,}~j} \left({\rm Swish} \left(FC_{\text{gate},~j} \left(\cdot \right) \right)\odot FC_{\text{up,}~j} \left(\cdot \right)  \right)}, \\
     \mathrm{FC_{i,j}}(\mathbf{x}_t) &= \mathrm{W_{i,j}^t} \mathbf{x}_t + \mathrm{b_{i,j}^t}, ~ i \in \{\text{up},\text{gate},\text{down}\}
\end{aligned}
\end{equation}

where 
$\{\mathrm{W}_{\text{up}}^{t},\mathrm{W}_{\text{gate}}^{t}\} \in \mathbb{R}^{D \times nD/E}$, 
$\{\mathrm{b}_{\text{up}}^{t},\mathrm{b}_{\text{gate}}^{t}\} \in \mathbb{R}^{nD/E}$,
$\mathrm{W}_{\text{down}}^{t} \in \mathbb{R}^{nD/E \times D}$, 
$\mathrm{b}_{\text{down}}^{t} \in \mathbb{R}^{D}$. $E$ is the expert number of each token. $k$ is the top-k number for the router $g(\cdot)$. And $g(\cdot)$ of the selected experts is constrained to a sum of \textbf{1} by ${\rm softmax(\cdot)}$, representing that the sum of the probabilities of activated experts is \textbf{1}. Detailed design concept and experimental results can be found in Appendix~\ref{sec:first_enlarge_then_sparse}.


\subsubsection{Shared Expert} Currently, several LLMs based on MoE architectures mention the use of shared experts to improve training stability and effectiveness, a design we have also adopted in practice. Notably, here the shared expert is still assigned per token, rather than being a globally visible shared expert for all tokens.
\begin{equation}
\begin{aligned}
\mathrm{S\textbf{-}P~MoE(\cdot)} =\sum_{i=1}^{k-1} g_{i}(\cdot)\cdot \text{Expert}_{i}(\cdot) + \text{SharedExpert}_{i}(\cdot)
\end{aligned}
\end{equation}

\subsubsection{Gate Value Scaling}
In practice, we observed that since $g(\cdot)$ typically uses softmax to convert router logits into a probability distribution summing to $\textbf{1}$, this sum-to-one constraint can lead to insufficient gradient updates for the SwiGLUs. Therefore, we add a constant $\mathbf{\alpha}$ before $g(\cdot)$ to allow more sufficient updates of the expert networks:
\begin{equation} \begin{aligned} \mathrm{S\textbf{-}P~MoE(\cdot)} = \mathbf{\alpha} \cdot  \sum_{i=1}^{k-1} g_{i}(\cdot)\cdot \text{Expert}_{i}(\cdot) + \text{SharedExpert}_{i}(\cdot) 
\end{aligned}
\end{equation}

\begin{table*}[!htbp]
\centering
\vspace{-0.3cm}
\caption{The time consumption, proportion, and bottlenecks of the main operators in TokenMixer-Large block.}
\vspace{-0.3cm}
\label{tab:operators_statics}
\scalebox{0.8}{
\begin{tabular}{c|ccc|ccc}
\hline
\multirow{2}{*}{Operator} & \multicolumn{3}{c|}{Training}         & \multicolumn{3}{c}{Serving}          \\ \cline{2-7} 
                          & Time(ms) & Percentage & Type          & Time(ms) & Percentage & Type         \\ \hline
MoEGroupedFFN             & 136.77   & 89.18\%    & Compute Bound & 7.43     & 98.35\%    & Memory Bound \\
MoEPermute                & 6.32     & 4.12\%     & Memory Bound  & 0.06     & 0.75\%     & Memory Bound \\
MoEUnpermute              & 10.27    & 6.69\%     & Memory Bound  & 0.07     & 0.90\%     & Memory Bound \\ \hline
\end{tabular}
}
\end{table*}

\subsubsection{Down-Matrix Small Init}
Inspired by Rezero~\cite{ramesh2021zero}, we reduce the initialization variance of the final layer W\_down in SwiGLU (we use \emph{xavier\_uniform} and lower the standard deviation hyperparameter from default 1 to 0.01). This adjustment keeps the $\mathrm{F(x)}$ term in $\mathrm{F(x)+x}$ close to zero during the early training phase, effectively making the module behave like an approximate identity mapping initially, thereby improving training stability. It also helps alleviate the output value explosion issue caused by the multiplicative interaction between the up and gate matrices in SwiGLU.

With the above modifications combined, our MoE model version can achieve performance nearly matching that of a fully activated dense model while activating only half of the parameters. We also observe from Figure~\ref{fig:load_balance} that after introducing these changes, the load is well balanced across tokens, eliminating the need for an additional load-balancing loss. Detailed analysis can be found in Appendix ~\ref{sec:load_balance}


\begin{figure}
    \centering
    \includegraphics[width=\linewidth]{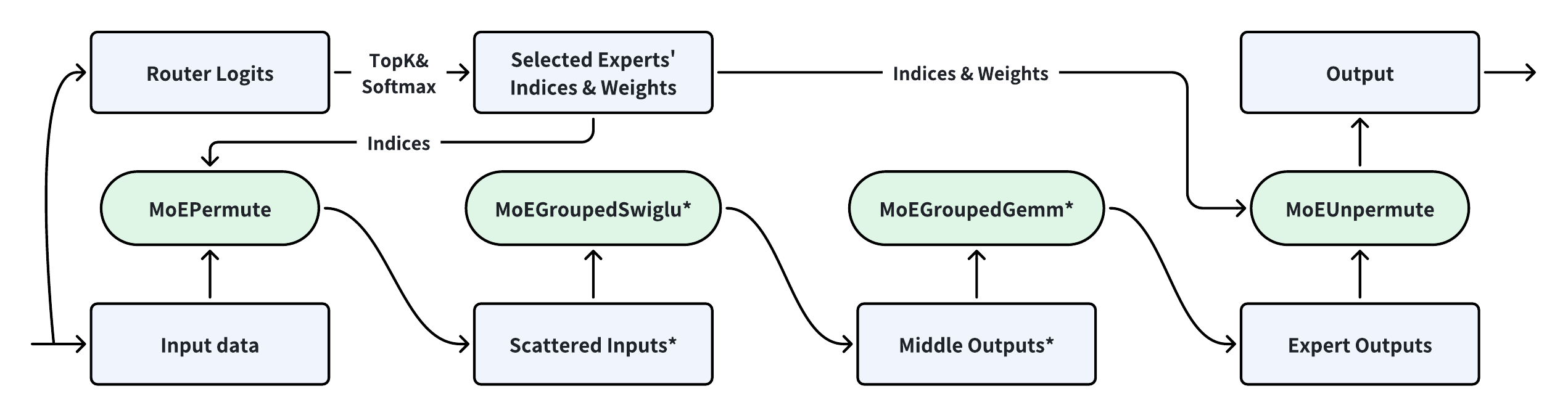}
    \vspace{-0.3cm}
    \captionsetup{skip=-2pt}
    \caption{Workflow of high-performance operators in one block. Green nodes represent operators, blue nodes represent data. The asterisk (*) indicates that the data is stored and computed in FP8 quantization.}
    \label{fig:operators_workflow}
\end{figure}
\subsection{Training/Serving Optimization}


We have developed a set of high-performance operators to enhance the efficiency of TokenMixer-Large in both training and inference.

\subsubsection{High Performance custom operators}
The overall workflow is illustrated in figure \ref{fig:operators_workflow}. MoEPermute transforms the input from batch-first to expert-first, ensuring that the input for each expert FFN is contiguous. Subsequently, the MoEGroupedFFN operators~(MoEGroupedSwiglu and MoEGroupedGemm)~compute all expert FFNs using a single kernel, reducing operator scheduling overhead and improving device utilization. Finally, MoEUnpermute calculates the weighted sum of the outputs from multiple activated experts as the final output.
The time consumption of each operator during training and inference is shown in Table \ref{tab:operators_statics}, with MoEGroupedFFN being the primary contributor to latency. Its proportion of inference time is even higher because its memory access does not decrease with smaller batch sizes during inference.

\subsubsection{FP8 Quantization}
We employ FP8 E4M3 post-training quantization during inference, while maintaining bfloat16 precision throughout training. Online experiments demonstrate that FP8 serving provides a 1.7x speedup without compromising model accuracy. 
As shown in Figure \ref{fig:operators_workflow}, the input is quantized in MoEPermute, and all expert weights have been pre-quantized.
Next, MoEGroupedSwiglu fuses the quantization of outputs and directly produces quantized activations.
Similarly, MoEGroupedGemm performs computations using FP8 and outputs results in bfloat16 format.

\subsubsection{Token Parallel}
Scaling the TokenMixer-Large architecture in distributed multi-device environments introduces significant bottlenecks. For \textbf{training}, the primary challenges are GPU memory capacity and overall training time. As models grow, their parameters and activations can exceed the memory of a single device. For \textbf{serving}, the issue is even worse: online inference with small batch sizes makes the computation memory-bandwidth-bound, leading to poor hardware utilization and low throughput. A naive model-parallel solution would shard the pertoken operation weights to each device, requiring an 'all2all' operation both before and after each computation to switch sharding layouts. This results in a total of four communication steps per block. To address these problems, we present \textbf{Token Parallel}, a model-parallelism strategy specifically designed for the TokenMixer-Large architecture. It partitions the model's parameters (specifically the pertoken operation weights) and computation across multiple devices.

\paragraph{Implementation}
The implementation of Token Parallel aligns the data sharding with the computation flow of the TokenMixer-Large block, which consists of a "Mixing" and a "Reverting" step, followed by a pertoken operation. All tensors below can be viewed as distributed tensors. We use \texttt{Shard(dim)} to denote a tensor sharded on a specific dimension.

\begin{lstlisting}[style=tfstyle, caption={Pseudocode of Token Parallel Data Flow.}, label={alg:token_parallel}]
# Input: X_prev [N*B, T, D]. Layout: Shard(Batch)
# N: Number of devices.
# B: local bsz

# Mixing
X_chunk = split(X_prev, num_splits=N, axis=2) # Shape: H x [N*B, T, D/H], Shard(0)
X_head = concat(X_chunk, axis=0).view(H, N*B, -1) # Shape: [H, N*B, T*D/H], Shard(1)

# Exchange data. After this, we hold full 'Batch' but sharded 'Head'
X_hidden = all2all(X_head) # Shape: [H, N*B, T*D/H], Shard(0)
h_hat = pertokenOperation(X_hidden) # Shape: [H, N*B, T*D/H], Shard(0)

# Reverting
chunks_t = split(h_hat, num_splits=T, axis=2).transpose(0, 1) # Shape: T x [N*B, H, D/H], Shard(1)

# Pack for transmission & merge Head dim
payload_t = concat(chunks_t).view(T, N*B, -1) # Shape: [T, N*B, D], Shard(2)

# Exchange data. After this, we hold full 'Dim' but sharded 'Token'
x_revert = all2all(payload_t) # Shape: [T, N*B, D], Shard(0)
x_hat = pertokenOperation(x_revert) # Shape: [T, N*B, D], Shard(0)

# 3. Final Output
X_curr = transpose(x_hat) # Shape: [N*B, T, D], Shard(1)
\end{lstlisting}

With Token Parallel, the output $\mathbf{X}_{l}$ remains sharded by the token dimension (\texttt{Shard(token)}) and is fed directly to the next block, whose initial 'split' operation can work on this layout. This eliminates the two extra communication steps, reducing the overhead from $4L$ to $2L+1$ for $L$ layers, as a final \texttt{all\_to\_all} is needed only at the end to restore the batch sharding.

The Token Parallel strategy delivers substantial performance gains across both training and inference. In production serving, a TokenMixer-Large model with 4-way token parallelism (global batch size 320) achieves a \textbf{29.2\%} throughput improvement over a non-parallelized baseline. This gain is further boosted to an impressive \textbf{96.6\%} by overlapping communication with computation, employing techniques like fine-grained micro-batch scheduling or in-kernel computation-communication overlapping. 

%% file: text/experiment.tex
\begin{figure*}[htbp]
    \centering
    \vspace{-0.2cm}
    \begin{tabular}{ccc}
        \includegraphics[width=0.33\textwidth]{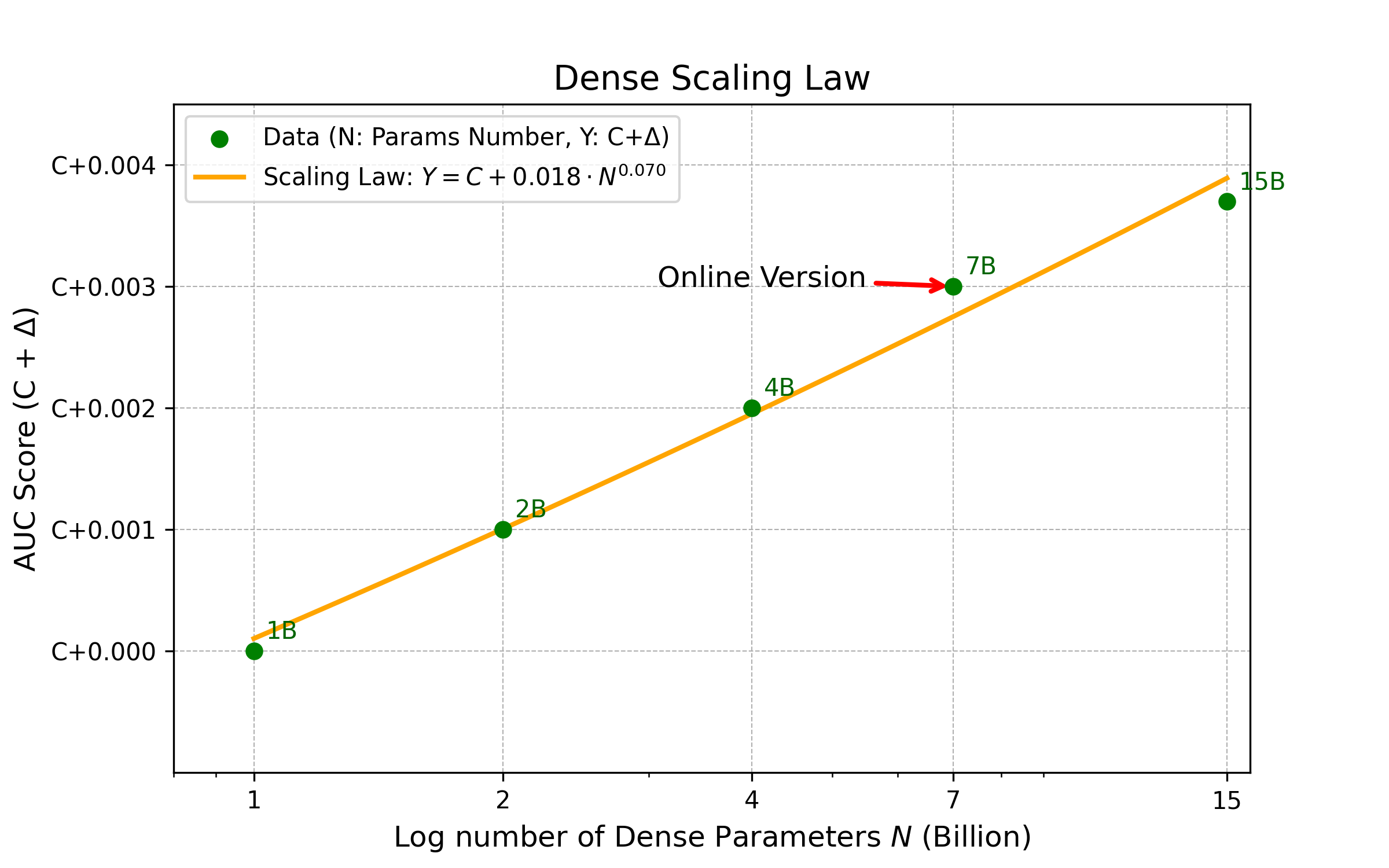} &
        \includegraphics[width=0.33\textwidth]{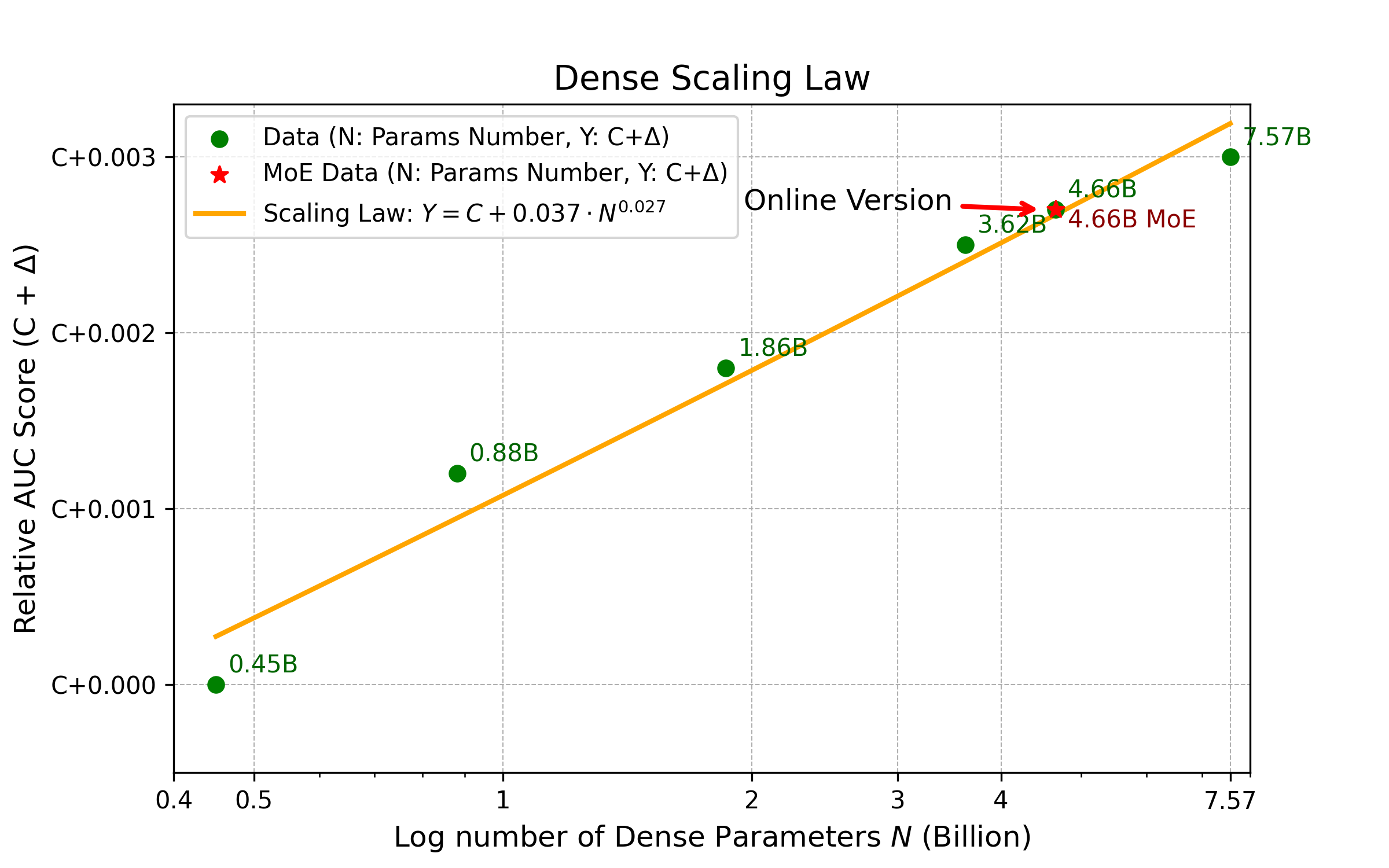} &
        \includegraphics[width=0.33\textwidth]{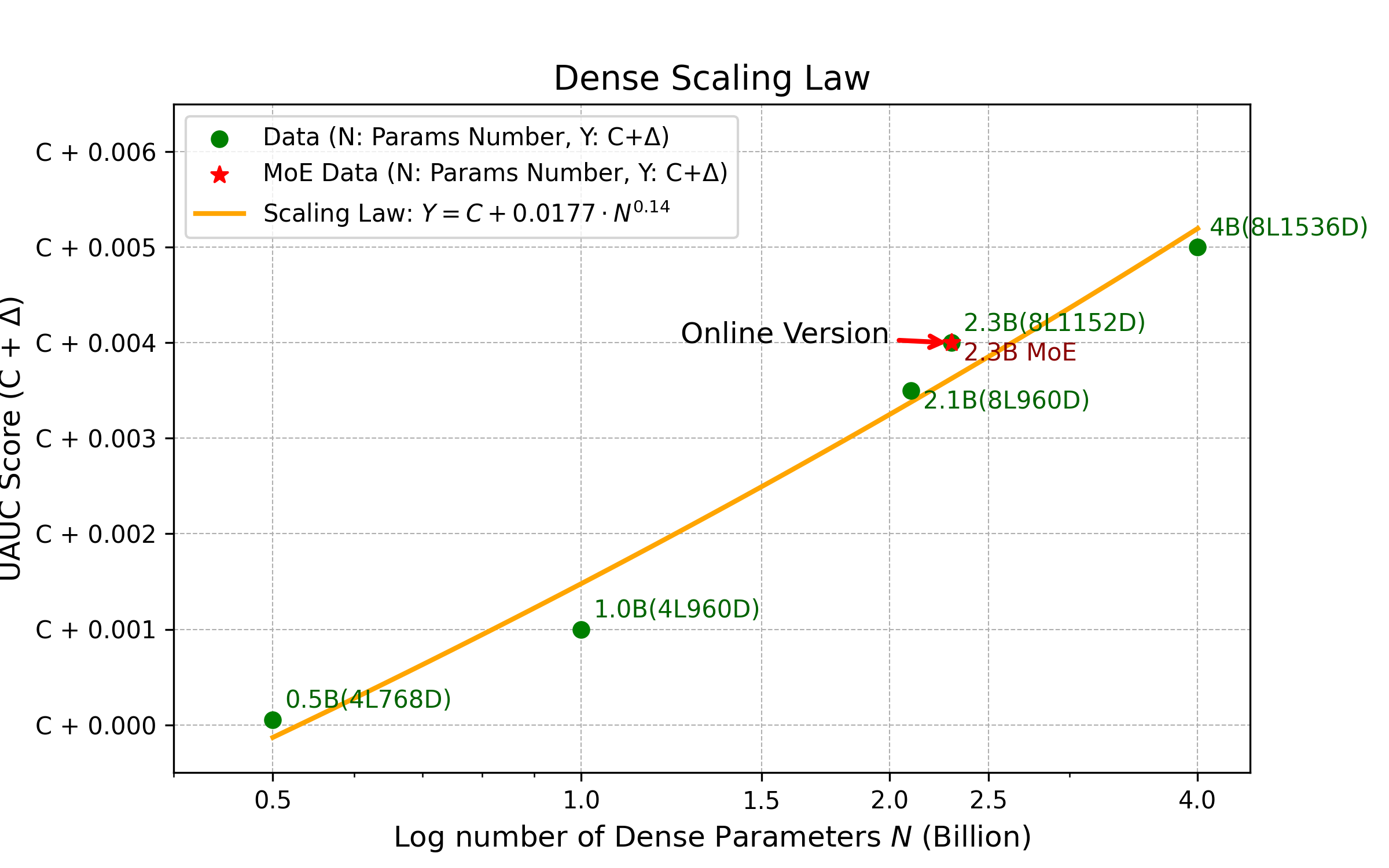} \\
        (a) Feed Ads 15B & (b) E-Commerce 7B & (c) Live Streaming 4B
    \end{tabular}
    \vspace{-0.2cm}
    \caption{Scaling Laws on different scenarios}
    \vspace{-0.2cm}
    \label{fig:Scaling_law_all}
\end{figure*}
\vspace{-0.2cm}

\section{Experiments}
\label{sec:experiments}
\subsection{Experiment Settings}
\subsubsection{Datasets and Environment}
The offline experiments in this study were mainly conducted on a real-world training dataset from the E-Commerce scenario on Douyin's main feed. This dataset is derived from the platform's online logs and user feedback labels, which include metrics such as product clicks, conversions, and Gross Merchandise Volume (GMV). The training set comprises over 500 distinct features—including numerical, ID-based, cross, and sequential features—and covers hundreds of millions of unique users. After a sampling process, the dataset contains approximately 400 million records per day, collected over a two-year period.

 Offline experiments also include real-world training data from Douyin Ads and Douyin Live Streaming. After sampling, the daily volumes reach 300 million and 17 billion records, respectively.


\subsubsection{Evaluation Metrics}

We use AUC~(Area Under the Curve) and UAUC~(User-Level AUC) as the primary effectiveness metrics, and dense parameter count, FLOPs, and MFU as efficiency metrics, detailed as follows: \textbf{AUC/UAUC}: We employ AUC from CTR and CVR tasks as the main effectiveness indicators.
\textbf{Parameters}: We use the number of pure dense parameters (excluding sparse embeddings) as one of the efficiency measures.
\textbf{Training FLOPs per Batch}: Refers to the number of floating-point operations required for the model to process a single batch of 2048 samples, representing the computational cost of training.
\textbf{MFU}:Model FLOPs Utilization Measures the actual utilization efficiency of the hardware’s theoretical floating-point computing capacity by the model.

\subsubsection{Baselines}
We mainly compare the following widely recognized SOTA baseline models: \textbf{DLRM-MLP}~\cite{rumelhart1986learning}, which employs multi-layer perceptrons (MLP) for feature interaction. \textbf{DCNv2}~\cite{wang2021dcn}, an improved version of "Deep \& Cross Network" that provides deeper and stronger feature-crossing capability. \textbf{AutoInt}~\cite{autoint}, which aims to automatically and explicitly learn high-order feature combinations through multi-head self-attention. \textbf{HiFormer}~\cite{gui2023hiformer}, which modifies the standard self-attention mechanism by using heterogeneous self-attention layers and low‑rank approximation. \textbf{DHEN}~\cite{zhang2022dhen}, which adopts a bagging idea to integrate multiple types of feature interaction modules (DCN/SA/FMB) within one block and stacks multiple layers. \textbf{Wukong}~\cite{zhangwukong}, which bags a factorization machine module (FMB) and a linear compression module (LCB) into two modules, stacks multiple layers, and studies the scaling laws of feature interactions. \textbf{Group Transformer}: Our in‑house Transformer variant designed for modeling heterogeneous features. It reduces the computational complexity of self‑attention by grouping features and aggregating them into a small number of tokens,  and employs pertoken Q/K/V/O to enhance the model’s capacity for representing heterogeneous features.
\textbf{FAT}~\cite{yan2025scaling}: An upgraded version of Group Transformer, it uses orthogonal basis combinations to improve model performance based on pertoken Q/K/V/O.
\textbf{RankMixer}~\cite{zhu2025rankmixer}: Built around a TokenMixer block that replaces self‑attention, substantially lowering computational cost. 

All experiments are conducted in a hybrid distributed training framework composed of 64 GPUs in E-commerce (256 in Feed-Ads and Live-Streaming). The framework adopts asynchronous updates for sparse parameters and synchronous updates for dense parameters. The optimizer hyperparameters remain consistent across all models: both the dense and sparse parts use the Adagrad optimizer, with learning rates set to 0.01 and 0.05 respectively.

\input{table/main_result_v2}
\subsection{Comparison With SOTA Methods}
The results from Table~\ref{table:main_result} demonstrate that TokenMixer‑Large outperforms all SOTA models, achieving a +1.14\% AUC gain in CTCVR compared to the MLP-500M baseline. Notably, models such as Wukong, Group Transformer, and RankMixer employ a pertoken network design, which leads to significantly lower FLOPs (with a batch size of 2048) than other baselines. Moreover, we find that the Sparse‑Pertoken MoE can match the performance of its dense counterpart while activating only half of the parameters, substantially improving the training/inference ROI of the model.

\definecolor{teal_check}{RGB}{0, 128, 128}
\definecolor{crimson_cross}{RGB}{176, 23, 56}
\newcommand{\cmark}{\textcolor{teal_check}{\ding{52}}} 
\newcommand{\xmark}{\textcolor{crimson_cross}{\ding{56}}} 
\definecolor{green-check}{HTML}{228B22}  
\definecolor{red-cross}{HTML}{DC143C}    

\begin{table}[!htbp]
  \centering
  \caption{Detailed Comparison With Rankmixer(TokenMixer)}
  \label{tab:rankmixer_detailed}
  \resizebox{\linewidth}{!}{%
  \renewcommand{\arraystretch}{1.0}  

  \begin{tabular}{
    l                   
    l                   
    c                   
    c                   
    c                   
    >{\centering\arraybackslash}m{1.2cm}  
    >{\raggedleft\arraybackslash}m{1cm}   
    >{\raggedleft\arraybackslash}m{1cm}   
  }
    \toprule  
    \textbf{Model Version}  & \textbf{SR} & \textbf{OTR} & \textbf{TSA} & \textbf{AUC$\uparrow$} & \textbf{Params} & \textbf{FLOPs} \\
    \midrule  
    Group Transformer & \textcolor{green-check}{\checkmark} & \textcolor{green-check}{\checkmark} & \textcolor{green-check}{\checkmark} & -- & 500M & 4.5T \\
    \addlinespace[0.25em]  
    RankMixer w/o SR\&OTR &  \textcolor{red-cross}{\xmark} & \textcolor{red-cross}{\xmark} & \textcolor{red-cross}{\xmark} & {-0.20\%} & 510M & 4.2T \\
    \addlinespace[0.25em] 
    RankMixer w/o OTR &  \textcolor{green-check}{\checkmark} & \textcolor{red-cross}{\xmark} & \textcolor{red-cross}{\xmark} & {-0.13\%} & 510M & 4.2T \\
    \addlinespace[0.25em] 
    RankMixer & \textcolor{green-check}{\checkmark} & \textcolor{green-check}{\checkmark} & \textcolor{red-cross}{\xmark} & {+0.03\%} & 567M & 4.6T \\
    \addlinespace[0.25em]
    TokenMixer-Large & \textcolor{green-check}{\checkmark} & \textcolor{green-check}{\checkmark} & \textcolor{green-check}{\checkmark} & {+0.13\%} & 500M & 4.2T \\
    \bottomrule  
  \end{tabular}
}
\end{table}

\subsection{Detailed Comparison With Rankmixer(TokenMixer)}
\label{sec:Detailed_Comparison_With_Rankmixer}
We conducted a detailed comparison between TokenMixer‑Large and Rankmixer(TokenMixer). Three key design aspects are defined as follows:

\begin{itemize}[topsep=0.5pt, leftmargin=*]
    \item \textbf{Standard Residual~(SR)}: Indicates whether standard residual connections exist between blocks. 
    \item \textbf{Original Token Residual~(OTR)}: Indicates whether the semantic information from the original tokens is preserved and propagated to subsequent layers. For example, if an input of 40 tokens is mixed into 16 new tokens and remains at a size of 16 tokens thereafter, the original semantic information of the 40 tokens cannot be retained all the way to the final output.
    \item \textbf{Token Semantic Alignment in Residual~(TSA)}: Refers to whether the semantic meaning of tokens remains consistent before and after the residual operation $F(x') + x$. In RankMixer, $x'$ in $F(x')$ corresponds to tokens after mixing, while the residual input $x$ comes from tokens before mixing, leading to a semantic mismatch.
\end{itemize}
Since each layer of Group Transformer does not change the number of input tokens, and the semantics of each position in \texttt{token\_dim} remain aligned after computing attention and pertokenFFN, it satisfies all three properties mentioned above. In contrast, RankMixer directly adds tokens before and after mixing in its residual connections, which breaks the semantic alignment across positions in \texttt{token\_dim}. Moreover, unless the number of new tokens is deliberately designed to match the original token count in RankMixer, the original token residual cannot propagate effectively through the network. As can be seen in Table~\ref{tab:rankmixer_detailed}, TokenMixer‑Large precisely addresses these issues in RankMixer, satisfying all three design properties while achieving the best performance.

\begin{figure}[htbp]
    \centering
    \includegraphics[width=\linewidth]{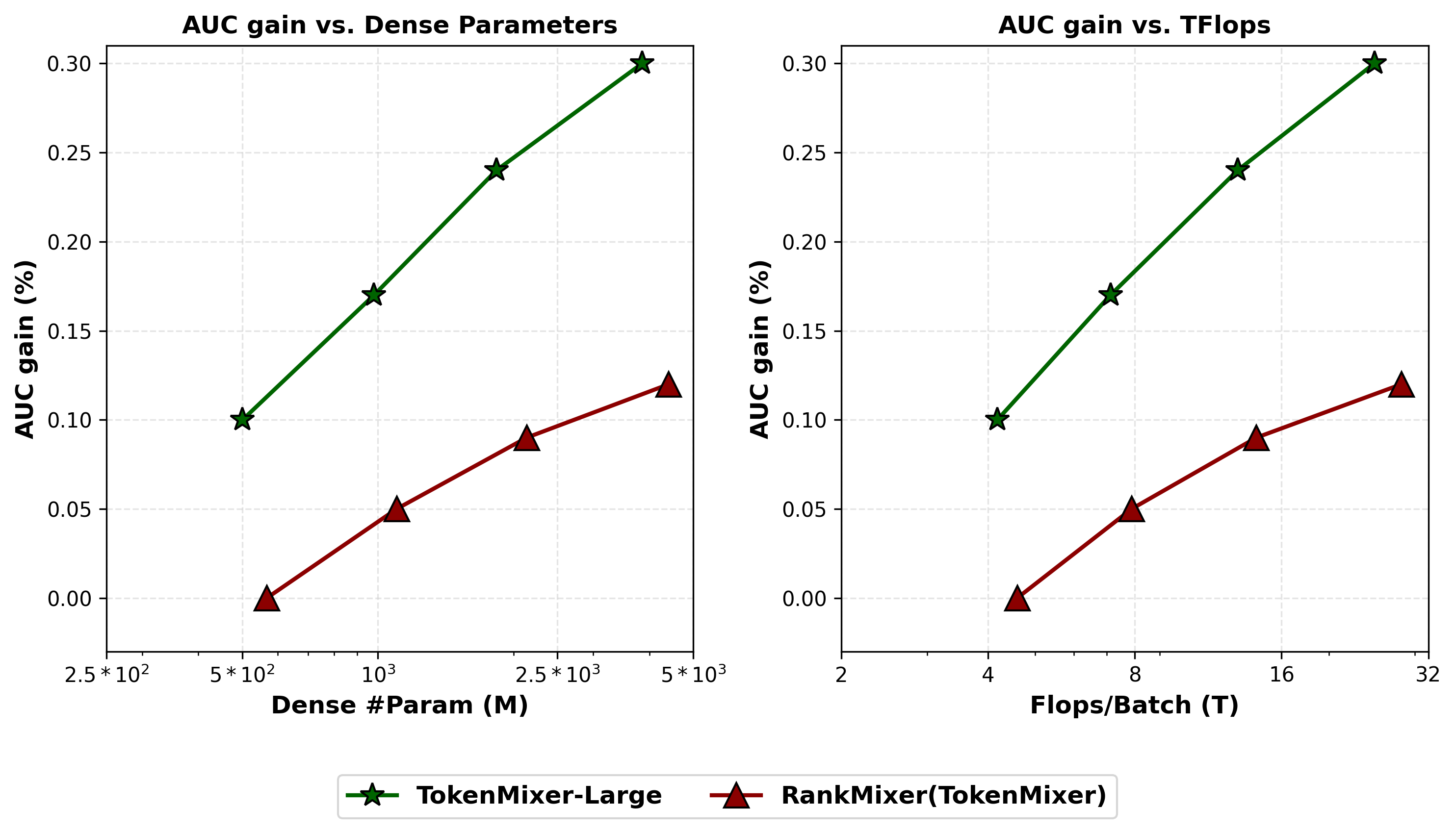}
    \vspace{-0.5cm}
    \caption{Scaling laws between AUC-gain and Params/Flops of various SOTA models. The x-axis uses a logarithmic scale.}
    \vspace{-0.5cm}
    \label{fig:scaling_law_sota_models}
\end{figure}

\subsection{Scaling Laws}
\subsubsection{Scaling Laws of SOTA Models} In Figure~\ref{fig:scaling_law_sota_models}, we selected the two best-performing models— RankMixer and TokenMixer‑Large, and examined their scaling laws. The results show that the AUC of both two models increases with the growth of parameters/FLOPs. Overall, TokenMixer‑Large achieves the best performance and exhibits a steeper improvement slope.

\subsubsection{Scaling Laws of TokenMixer-Large on Different Scenarios}
We scale up the model by increasing its dimension $D$, depth $L$, and scaling factor $N$~(for SwiGLU hidden expansion). Two key findings emerge:
\begin{itemize}[topsep=0.5pt, leftmargin=*]
    \item \textbf{Balanced expansion across dimensions yields greater returns}: Unlike earlier RankMixer experiments below the 1B scale, we find that expanding any single dimension—width, depth, or the scaling factor—brings gains but gradually encounters a bottleneck. When scaling beyond 1B parameters, a balanced increase across all dimensions is needed to achieve better returns.
    \item \textbf{Larger models require more data for convergence}: As illustrated in Table~\ref{tab:model_convergence}, a model scaling from 30M to 90M parameters converges with only 14 days of training samples, whereas scaling from 500M to 2B requires 60 days of samples to converge fully.
\end{itemize}
We have also validated the scaling laws of TokenMixer-Large across different business scenarios, observing significant scaling law behavior in each case. Ultimately, in the Feed Ads, E-Commerce, and Live Streaming scenarios, we successfully scaled the model to \textbf{15B}, \textbf{7B} and \textbf{4B} parameters on offline experiments, as well as \textbf{7B}, \textbf{4B} and \textbf{2B} parameters on online traffic, respectively, achieving substantial scaling gains.

\definecolor{green-positive}{HTML}{228B22}  
\definecolor{red-negative}{HTML}{DC143C}    

    

\begin{table}[htbp]
  \centering
  \begin{threeparttable}
    \caption{Model Convergence Comparison on Douyin Live Streaming}
    \label{tab:model_convergence}
    \begin{tabular}{lcc} 
      \toprule
      \textbf{Param}\tnote{*} & \textbf{Convergence Day} & \textbf{$\Delta$UAUC} \\
      \midrule
      30m                     & --                       & --                        \\
      \midrule
      90m↑                    & 14d                      & +0.94\%                   \\
      \midrule
      500m↑                   & 30d                      & +0.62\%                   \\
      \midrule
      2.3B↑                   & 30d                      & +0.41\%                   \\
      2.3B↑ (60d)             & 60d                      & +0.70\%                   \\
      \bottomrule
    \end{tabular}
    \begin{tablenotes}[para,flushleft] 
      * Each row uses the \textit{previous row} as the baseline. The symbol ↑ indicates comparison against the prior parameter scale.
    \end{tablenotes}
  \end{threeparttable}
\end{table}




\subsection{Ablation Study}
\subsubsection{Ablation Study for TokenMixer-Large Block}
As can be seen from Table~\ref{table:ablation_study}, we conducted ablation studies on different modules in TokenMixer-Large, including Tokenization, mixing\&reverting, pertoken SwiGLU, and residual connections. Results show that removing any of these modules leads to performance degradation. Among them, "mixing \& reverting" and Pertoken-SwiGLU have the most significant impact on overall performance.
\paragraph{Pure Model Design:}
With the continuous expansion of the number of parameters in tokenmxier-large, the benefits of these numerous small, I/O-bound, fragmented operators~(Such as DCN~\cite{wang2021dcn} and LHUC~\cite{swietojanski2016learning}) can all be captured by tokenmxier-large itself. Furthermore, since tokenmixer-large itself \textbf{only contains} parameterless mixing and reverting operations as well as a large number of GroupedGemm operations, it is highly efficient. All detailed analysis can be found in Appendix ~\ref{sec:Pure_Model_Design}

\paragraph{Normalization:} 
We conducted a thorough analysis of the positions for applying normalization, including pre-norm, post-norm, and sandwich norm. All detailed analysis can be found in Appendix ~\ref{sec:Normalization_Position}

\paragraph{Mixing Strategy:} Finally, we conducted a detailed study on the splitting strategy of the mixer. All detailed analysis can be found in Appendix ~\ref{sec:Mixing_Strategy}

\begin{table}[t]
\centering
\caption{Ablation on components of TokenMixer-Large 4B}
\vspace{-0.3cm}
\label{table:ablation_study}
\small
\begin{tabular}{@{\quad}l c@{\quad}}
\toprule
\textbf{Setting} & $\Delta$AUC \\ 
\midrule
w/o Global Token                         & $-0.02\%$ \\
w/o Mixing \& Reverting                 & $-0.27\%$ \\
w/o Residual              & $-0.15\%$ \\
w/o Internal Residual \& AuxLoss        & $-0.04\%$ \\
Pertoken SwiGLU   $\rightarrow$ SwiGLU       & $-0.21\%$ \\
Pertoken SwiGLU   $\rightarrow$ Pertoken FFN       & $-0.10\%$ \\
\bottomrule
\end{tabular}
\end{table}


\begin{table}[H]
\centering
\caption{Ablation on components of Sparse-Pertoken MoE}
\vspace{-0.3cm}
\label{table:MoE_ablation_study}
\small
\begin{tabular}{@{\quad}l c c c@{\quad}}
\toprule
\textbf{Setting} & $\Delta$AUC & $\Delta$Params & $\Delta$FLOPs \\ 
\midrule
w/o Shared Expert & $-0.02\%$ & $ 0.0$\% & $ 0.0\%$ \\
w/o Gate Value Scaling      & $-0.03\%$ & $ 0.0$\% & $0.0$\% \\
w/o Down-Matrix Small Init & $-0.03\%$ &  0.0\% & 0.0\% \\
Sparse-Pertoken MoE $\rightarrow$ Sparse MoE & $-0.10\%$ &  0.0\% & 0.0\% \\
\bottomrule
\end{tabular}
\end{table}

\subsubsection{Ablation Study for Sparse-Pertoken MoE}
As shown in Table~\ref{table:MoE_ablation_study}, We ablated different modules in the Sparse-Pertoken MoE, including the shared expert, Gate Value Scaling and Down-Matrix Small Init, we found that each of them contributes positively to the final performance improvement. The aforementioned operations introduce no additional parameters or computational overhead (FLOPs), and are therefore cost-free modifications.

Additionally, under the condition of aligning both the total parameter count and the activated parameter count, we compared the sparse-pertoken MoE with the standard MoE and observed a significant performance drop with the standard MoE. This is because sparse-pertoken MoE is equivalent to giving standard MoE a routing prior, meaning that the experts that each token can activate are not shared, thus avoiding the problem of routers having difficulty learning in the early stages.

\paragraph{Gate Value Scaling:} 
Meanwhile, we conducted comprehensive experiments on Gate Value Scaling and found that the optimal scaling factor varies for models with different sparsity ratios (activated parameters/total parameters). A model with a sparsity ratio of 1:2 performs best with a scaling factor of 2, whereas a model with a sparsity ratio of 1:4 achieves optimal results with a scaling factor of 4. The scaling factor exhibits an \textbf{inversely proportional} relationship with the sparsity ratio. Detailed analysis can be found in Appendix ~\ref{sec:Gate_Value_Scaling_Analysis}.

\paragraph{Down Matrix Small Initialization:} 
Initializing the last down matrix $\mathbf{FC_{down}}$ in SwiGLU to 0.01 times that of the $\mathbf{FC_{up}/FC_{gate}}$ layers can significantly improve the model's performance and training stability. Detailed analysis can be found in Appendix ~\ref{sec:Down_Matrix_Small_Initialization}.



\begin{table}[H]
\centering
\caption{Online performance of TokenMixer-Large}
\vspace{-0.3cm}
\label{table:online_performance}
\small
\begin{tabular}{c|cc|ccc|cc}
\toprule
    \multirow{2}{*}{} & \multicolumn{2}{c|}{Feed Ads} & \multicolumn{3}{c|}{E-Commerce} & \multicolumn{2}{c}{Live Streaming} \\
    \midrule
    & $\Delta$AUC & \textbf{ADSS} & $\Delta$AUC & \textbf{Order} & \textbf{GMV} & $\Delta$UAUC & \textbf{Pay} \\
    \midrule
    Lift↑  & 0.35\% & 2.0\% & 0.51\% & 1.66\% & 2.98\% & 0.7\% & 1.4\%  \\
\bottomrule
\end{tabular}
\end{table}

\subsection{Online Performance}

To validate the effectiveness of TokenMixer-Large, we have deployed it across multiple scenarios at Douyin, including feed advertising, live streaming e-commerce and general live streaming. These deployments cover a majority of Douyin's businesses and serve hundreds of millions of users. For each business scenario, we employ distinct Key Performance Indicators:
\begin{itemize}[leftmargin=10pt]
    \item \textbf{Douyin Live-Streaming E-Commerce}: Gross Merchandise Volume (GMV) and order count are used as the core online business metrics.
    \item \textbf{Douyin Feed Advertising}: Advertiser Satisfaction Score (ADSS) and Advertiser Value (ADVV) serve as the primary online business indicators.
    \item \textbf{Douyin Live Streaming (Non-E-Commerce)}: Total Payment Amount is adopted as the core online business metric.
\end{itemize}
\input{table/online_abtest}
For the advertising, e-commerce, and live-streaming scenarios, the online baselines are RankMixer-1B, RankMixer-150M, and RankMixer-500M, respectively. We scale up TokenMixer-Large to 7B, 4B, and 2B parameters correspondingly. As can be seen from Table~\ref{table:online_performance}, the results show that TokenMixer‑Large delivers an increase of +1.66\% in orders and +2.98\% in per‑capita preview payment GMV for e‑commerce; improves ADSS by +2.0\% in advertising; and achieves a +1.4\% revenue growth for live streaming.



%% file: table/main_result_v2.tex
\vspace{-10pt} 
\begin{table}[H]
\centering
\caption{Performance \& efficiency comparison of $\sim$500 M-parameter models on E-commerce (best values in \textbf{bold}).}
\vspace{-0.3cm}
\label{table:main_result}
\resizebox{\linewidth}{!}{%
\renewcommand\arraystretch{1.05}
\begin{tabular}{lccc}
\toprule
\multirow{2}{*}{Model} & \multicolumn{1}{c}{\textbf{CTCVR}} & \multicolumn{2}{c}{\textbf{Efficiency}} \\
\cmidrule(lr){2-2}  \cmidrule(l){3-4}
 & $\Delta$AUC$\uparrow$ & Params & FLOPs/Batch \\
\midrule
DLRM-MLP-500M~\cite{rumelhart1986learning}    & --  & 499 M  &  125.1 T \\

\midrule
HiFormer~\cite{gui2023hiformer}             & +0.44\% & 570 M & 28.8 T \\
DCNv2~\cite{wang2021dcn}                & +0.49\% & 502 M  & 125.8 T \\
DHEN~\cite{zhang2022dhen}                 & +0.63\% & 415 M  & 103.4 T \\
AutoInt~\cite{autoint}              & +0.75\% & 549 M & 138.6 T \\
Wukong~\cite{zhangwukong}               & +0.76\% & 513 M & 4.6 T \\
Group Transformer            & +0.81\% & 550 M & 4.5 T \\
FAT~\cite{yan2025scaling}   & +0.82\% & 551 M & 4.59 T \\
RankMixer(TokenMixer)~\cite{zhu2025rankmixer}         & +0.84\% & 567 M & 4.6 T \\

\midrule
\textbf{TokenMixer-Large 500M} & \textbf{+0.94\%} & 501 M & 4.2 T \\
\textbf{TokenMixer-Large 4B}   & \textbf{+1.14\%}  & \textbf{4.6 B} & \textbf{29.8 T} \\
\textbf{TokenMixer-Large 7B}   & \textbf{+1.20\%}  & \textbf{7.6 B} & \textbf{49.0 T} \\
\textbf{TokenMixer-Large 4B SP-MoE}   & \textbf{+1.14\%} & \textbf{2.3 B in 4.6 B} & \textbf{15.1 T} \\
\bottomrule
\end{tabular}%
}
\end{table}
\vspace{-10pt} 

%% file: table/online_abtest.tex
\begin{table*}[H]
\caption{Online AB test result for Feed Recommendation Scenarios in both Douyin and Douyin lite app and the improvements are all statistically significant. According to long-term reverse A/B tests and continuous observations over long-term reverse experiments, the gains provided by RankMixer-1B have not yet converged, and the results presented here are still improving.}
\vspace{-0.3cm}
\resizebox{\textwidth}{!}{
\renewcommand\arraystretch{1.08}
\begin{tabular}{lcccclccccl}
\hline
                       & \multicolumn{5}{c}{\textbf{Douyin app}}                                                      & \multicolumn{4}{c}{\textbf{Douyin lite}}                                  &                  \\ \cmidrule(lr){2-6}\cmidrule{7-11} 
                       & \textbf{Active Day}$\uparrow$ & \textbf{Duration}$\uparrow$ & \textbf{Like}$\uparrow$ & \textbf{Finish}$\uparrow$ & \textbf{Comment}$\uparrow$ & \textbf{Active Day}$\uparrow$ & \textbf{Duration}$\uparrow$ & \textbf{Like}$\uparrow$ & \textbf{Finish}$\uparrow$ & \textbf{Comment}$\uparrow$ \\ \hline

\textbf{Overall}       & +0.2908\%           & +1.0836\%         & +2.3852\%     & +1.9874\%       & +0.7886\%        & +0.1968\%           & +0.9869\%          & +1.1318\%     & +2.0744\%        & +1.1338\%        \\ \hline 
\textbf{Low-active}    & +1.7412\%            & +3.6434\%         & 
+8.1641\%     & +4.5393\%       & +2.9368\%         & +0.5389\%           & +2.1831\%         & +4.4486\%      & +3.3958\%       & +0.9595\%        \\
\textbf{Middle-active} & +0.7081\%           & +1.5269\%         & +2.5823\%     & +2.5062\%       & +1.2266\%         & +0.4235\%           & +1.9011\%         & +1.7491\%      & +2.6568\%         & +0.6782\%        \\
\textbf{High-active}   & +0.1445\%           & +0.6259\%         & +1.828\%      & +1.4939\%       & +0.4151\%          & +0.0929\%           & +1.1356\%         & +1.8212\%      & +1.7683\%        & +2.3683\% 
\\ \hline
\end{tabular}
}
\label{online_ab}
\end{table*}


%% file: text/conclusion.tex
\section{Conclusion}
\label{sec:conclusion}
In this paper, we present TokenMixer-Large, an upgraded version of TokenMixer. We conduct an in‑depth analysis of the limitations in the original TokenMixer design and provide a detailed investigation and discussion on deep layer modeling and Sparse‑Pertoken MoE. On the engineering side, we implement a set of efficient MoE operators and a Token‑Parallel strategy specifically designed for Pertoken experts. TokenMixer-Large has delivered significant offline and online gains across multiple scenarios at ByteDance, serving hundreds of millions of users.


%% file: text/appendix.tex


\begin{figure*}[!t]
    \centering
    \includegraphics[width=1.0\textwidth]{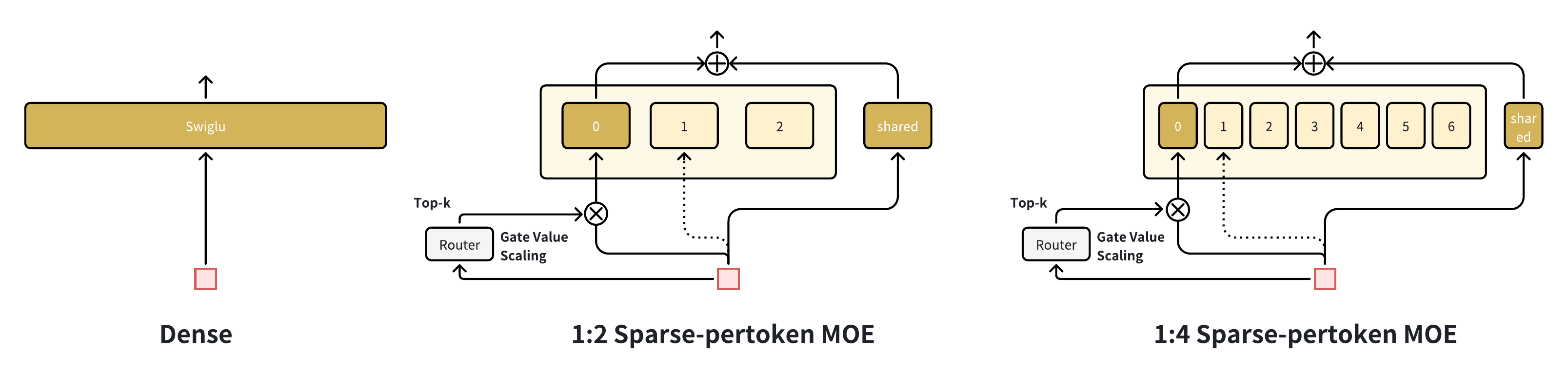}
    \captionsetup{skip=0pt}
    \caption{First Enlarge Then Sparse illustration}
    \label{fig:fisrt_enlarge_then_sparse}
\end{figure*}

\newpage
\appendix
\section{Appendix}
\subsection{First Enlarge Then Sparse}
\label{sec:first_enlarge_then_sparse}
The conventional approach to MoE is to directly increase the total number of parameters while keeping the number of activated experts constant. However, we believe that this method of increasing experts is equivalent to expanding the hidden expansion of Per-token SwiGLU in a single dimension, which always has an upper limit that is easily reached. Therefore, we choose to first design a dense model with the best performance, and then achieve efficiency gains by sparsifying it while minimizing performance degradation.

Therefore, we choose the "First Enlarge, Then Sparse" approach. Specifically, we will refine all the FC networks in SwiGLU within TokenMixer-large into fine-grained version and then perform sparse activation. Figure~\ref{fig:fisrt_enlarge_then_sparse} illustrates the specific forms of dense model, 1:2 sparsity MoE, and 1:4 sparsity MoE. Currently, we can achieve near-zero offline/online drop with a sparsity of 1:2, and a slight decrease with a sparsity of 1:4. However, as the sparsity increases, the GEMM component in SwiGLU will further decrease, so greater sparsity gains depend on a larger model and a higher GEMM ratio. \textbf{Therefore, we chose the version with the highest ROI of 1:2 for online deployment}. Whether we can achieve sparsity greater than 1:8 while maintaining performance without loss is still under exploration.

\subsection{Load Balance}
\label{sec:load_balance}


\begin{figure}[!htbp]
    \centering
    \includegraphics[width=1.0\linewidth]{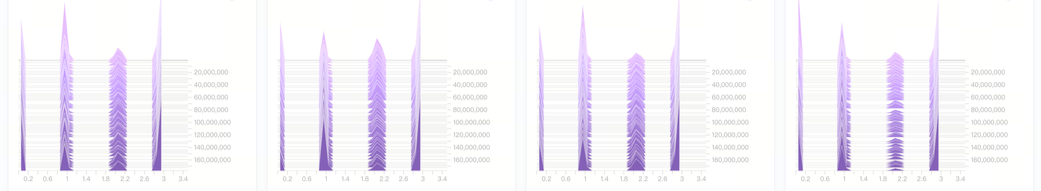} \\ 
    \vspace{5pt}
    \includegraphics[width=1.0\linewidth]{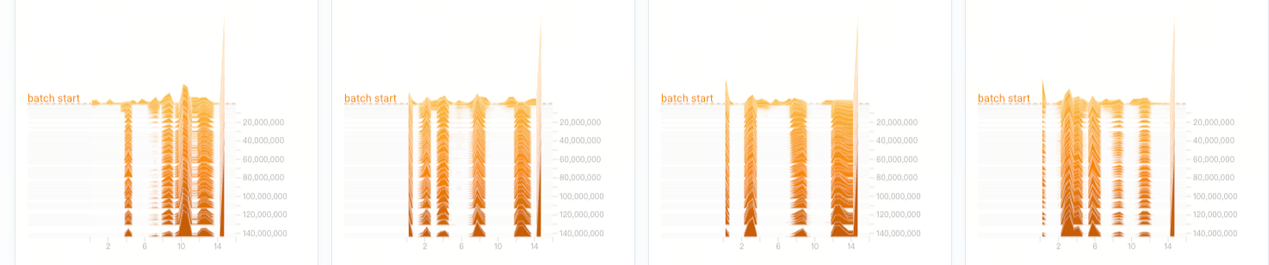}
    \caption{Sparse-Pertoken MoE Load Balance of 1:2 and 1:8 versions}
    \label{fig:load_balance}
\end{figure}

The top one of Figure~\ref{fig:load_balance} shows a load balancing example on e-commerce 4.6B Sparse-Pertoken MoE. We use one shared expert and three routing experts, activating a total of two experts. With a sparsity ratio of 1:2, it achieves the same offline/online performance as the dense model~(4.6B dense) with the same total number of parameters as MoE. As shown, we illustrate the activation status of each token, where indices 0-2 represent routing experts and index 3 represents the shared expert. We can see that the overall load is relatively balanced. Since index 3 is the always-active shared expert, we can observe that the activation level of index 3 is higher than that of indices 0-2.

However, as the sparsity increases (1:2 to 1:8), the below one of Figure~\ref{fig:load_balance} shows that the load balancing of the model deteriorates to some extent. We also tried the load balancing method of Switch Transformer~\cite{fedus2022switch} and Z-loss~\cite{zoph2022st}, which have some benefits. However, considering that the currently launched 1:2 sparsity version has the highest ROI and does not have a load balancing problem, this part will be explored in the future.

\subsection{Pure Model Design}
\label{sec:Pure_Model_Design}

\begin{table}[!htbp]
  \centering
    \caption{The effect gain brought by DCN on tokenmixer-large at different parameter scales. The DCN Gain is equal to the effect of tokenmixer-large with DCN minus the effect of tokenmixer-large without DCN.}
    \label{tab:DCN_Gain}
    \begin{tabular}{ll}
      \toprule
      \textbf{Params} & \textbf{DCN Gain} \\
      \midrule
      150M                  &  +0.09\%                \\
      500M                  &  +0.04\%                \\
      700M                  &  +0.00\%                \\
      \bottomrule
    \end{tabular}
\end{table}
In our baseline model, we manually filtered the pooling results of sequences and fed them into a DCN module for feature cross-pollination. The output of this module is used as additional tokens to supplement the input of the backbone network's TokenMixer. However, this design has one main problems. As the complexity of the backbone network TokenMixer continues to increase, the effect gain brought by the front-end DCN gradually diminishes. Experimental data shows in Table~\ref{tab:DCN_Gain} that on a model with 150M parameters, DCN can bring a +0.09\% gain; however, when the model parameters are expanded to 700M, removing the DCN structure no longer causes any effect loss.

In addition to DCN, we also experimented with integrating structures such as DHEN and LHUC with TokenMixer‑Large, using both parallel and serial configurations. Practical results demonstrate that as the parameter scale of TokenMixer‑Large expands, the gains provided by these small, numerous, and IO‑bound fragmented operators can be fully captured by TokenMixer‑Large itself. Furthermore, the TokenMixer‑Large Block itself achieves a high Model FLOPs Utilization (MFU) of up to \textbf{60\%}.

\begin{table}[!htbp]
\centering
\caption{Ablation on variants of Normalization}
\vspace{-0.3cm}
\label{Norm_ablation_study}
\small
\begin{tabular}{@{\quad}l c@{\quad}}
\toprule
\textbf{Setting} & $\Delta$AUC \\ 
\midrule
Pre-Norm                        & -- \\
Post-Norm                & $+0.01\%$ $\rightarrow$ \textbf{NaN} \\
Sandwich-Norm            & $-0.03\%$ \\
\bottomrule
\end{tabular}
\end{table}


\subsection{Normalization and Position}
\label{sec:Normalization_Position}
RMSNorm can be formulated as follows:
\begin{equation}
\begin{aligned}
\text{RMS}(x) = \sqrt{\frac{1}{d} \sum_{i=1}^{d} x_i^2 + \epsilon} \\
\bar{a}_i = \frac{x_i}{\text{RMS}(x)} \\
\text{RMSNorm}(x) = \bar{a}_i \cdot \gamma \\
\end{aligned}
\end{equation}

where $\gamma$ is the learnable parameter for normalization. Considering that the bias of the kernel matrix is mathematically redundant when the input passes through the normalization layer,
and inspired by Llama~\cite{chowdhery2023palm, touvron2023llama}, we remove all bias kernel matrices and replaced all Norm in TokenMixer-large from LayerNorm to RMSNorm. In our tests, we achieved \textbf{8.4\%} increase in end-to-end throughput while maintaining the same performance.
We conducted a thorough analysis of the positions for applying normalization, including pre-norm, post-norm, and sandwich norm. As can be seen in Table~\ref{Norm_ablation_study}, in terms of effectiveness, post-norm delivers better performance but is prone to gradient explosion. Pre-norm, while less effective than post-norm, ensures stable training of the model.

\begin{table}[!htbp]
\centering
\caption{Ablation on Mixing Strategy}
\vspace{-0.3cm}
\label{mixer_ablation_study}
\small
\begin{tabular}{@{\quad}l c@{\quad}}
\toprule
\textbf{Setting} & $\Delta$AUC \\ 
\midrule
Vertical division                & -- \\
Diagonal division               & $+0.00\%$ \\
Random division                  & $+0.00\%$ \\
Vertical division (half of raw tokens) & $-0.08\%$ \\
\bottomrule
\end{tabular}
\end{table}


\subsection{Mixing Strategy}
\label{sec:Mixing_Strategy}
We conducted a detailed study on the splitting strategy of the mixer. As can be seen in Table~\ref{mixer_ablation_study}, we found that different split-concat strategies do \textbf{not} affect performance, \textbf{as long as each newly mixed token contains all the original token information}. As a control, we attempted to mix only half of the original token information, which resulted in a significant performance drop.

\subsection{Gate Value Scaling Analysis}
\label{sec:Gate_Value_Scaling_Analysis}
\begin{table}[htbp]
  \centering
  \caption{Impact of Model Sparsity and $\alpha$ Settings on AUC Performance} %
  \label{tab:sparsity_auc}
  \begin{tabular}{@{}llcc@{}} 
    \toprule
    \textbf{Version} & \textbf{Sparsity} & \textbf{$\alpha$ Setting} & \textbf{AUC Performance} \\ 
    \midrule
    4B dense & -- & -- & -- \\
    \midrule
    \multirow{4}{*}{\makecell{2B in 4B MoE}} & \multirow{4}{*}{1:2} & 1 & -0.04\% \\
    & & \textbf{2} & \textbf{-0.00\%} \\
    & & 3 & -0.02\% \\
    & & 4 & -0.05\% \\
    \midrule
    \multirow{5}{*}{\makecell{1B in 4B  MoE}} & \multirow{5}{*}{1:4} & 1 & -0.07\% \\
    & & 2 & -0.05\% \\
    & & \textbf{4} & \textbf{-0.03\%} \\
    & & 6 & -0.04\% \\
    & & 8 & -0.05\% \\
    \bottomrule
  \end{tabular}
\end{table}

During the process of converting a dense model into a MoE architecture, we found that introducing a constant scaling factor \(\alpha\) before the router function \(g(x)\)—by amplifying the router's output to enhance gradient updates for the selected (routed) experts—significantly improves model performance. With the inclusion of a shared expert, the MoE formulation can be expressed as:

\begin{equation}
\begin{aligned}
y = \sum_{i=1}^{K-1} g(x)_i \cdot \text{Expert}_i(x) + \text{SharedExpert}(x)
\end{aligned}
\end{equation}

where \(g(x)\) represents the routing weights, \(\text{Expert}_i\) denotes the \(i\)-th expert network, and \(\text{SharedExpert}\) is the shared expert network.

Thus, the mathematical form of the Sparse-pertoken MoE is as follows:

\begin{equation}
\begin{aligned}
y = \alpha \cdot \sum_{i=1}^{K-1} g(x)_i \cdot \text{Expert}_i(x) + \text{SharedExpert}(x)
\end{aligned}
\end{equation}

In practice, we observed that the setting of the scaling factor $\alpha$ is crucial: values that are too large or too small can degrade performance. Further experiments revealed that the optimal value of $\alpha$ is closely related to the model’s sparsity. Specifically, setting $\alpha$ \textbf{inversely proportional} to the current sparsity ratio yields the best results, as shown in Table~\ref{tab:sparsity_auc}.

We argue that the \textbf{Gate Value Scaling} operation is effective, and the reason for setting $\alpha$ to the reciprocal of the sparsity ratio is as follows: the gradient updates for each fully connected (FC) layer in SwiGLU need to remain consistent with those in the original dense model. Our "first enlarge, then sparse" strategy splits a large kernel in SwiGLU into several smaller kernels—similar to the fine‑grained expert approach used in DeepSeek‑MoE~\cite{dai2024deepseekmoe}, but with the key difference that we do not increase the number of activated parameters~(topk). This means that as the model is divided more finely, the probability of each expert being updated decreases. Therefore, we need to amplify the gradient magnitude for an expert when it is activated, which is precisely what Gate Value Scaling achieves.

Another possibility is that Gate Value Scaling increases the overall weighted sum-pooling result of $\alpha \cdot \sum_{i=1}^{K-1} g(x)_i \cdot \text{Expert}_i(x)$, thus increasing the variance of the output at each layer of the model. Therefore, we tried directly increasing the initial variance of the Expert kernel to simulate expanding the output of $\alpha \cdot \sum_{i=1}^{K-1} g(x)_i \cdot \text{Expert}_i(x)$ from another perspective. As shown in the table~\ref{fig:Increase_Variance}, directly increasing the initial variance of the Expert kernel did not improve the performance; instead, it led to a significant degradation in results.

\begin{table}[!htbp]
\centering
\caption{Gate Value Scaling vs. Increase Variance}
\vspace{-0.3cm}
\label{fig:Increase_Variance}
\small
\begin{tabular}{@{\quad}l c@{\quad}}
\toprule
\textbf{Setting} & $\Delta$AUC \\ 
\midrule
Gate Value Scaling                       & -- \\
Increase Variance               & $-0.07\%$ \\
\bottomrule
\end{tabular}
\end{table}

\subsection{Down-Matrix Small Initialization}
\label{sec:Down_Matrix_Small_Initialization}
To address the convergence challenges associated with deeper models, in addition to incorporating standard residual connections, interval residual connections, and residual loss (resloss), we also adopted the concept from Rezero~\cite{bachlechner2021rezero}. 

\begin{figure}
    \centering
    \includegraphics[width=\linewidth]{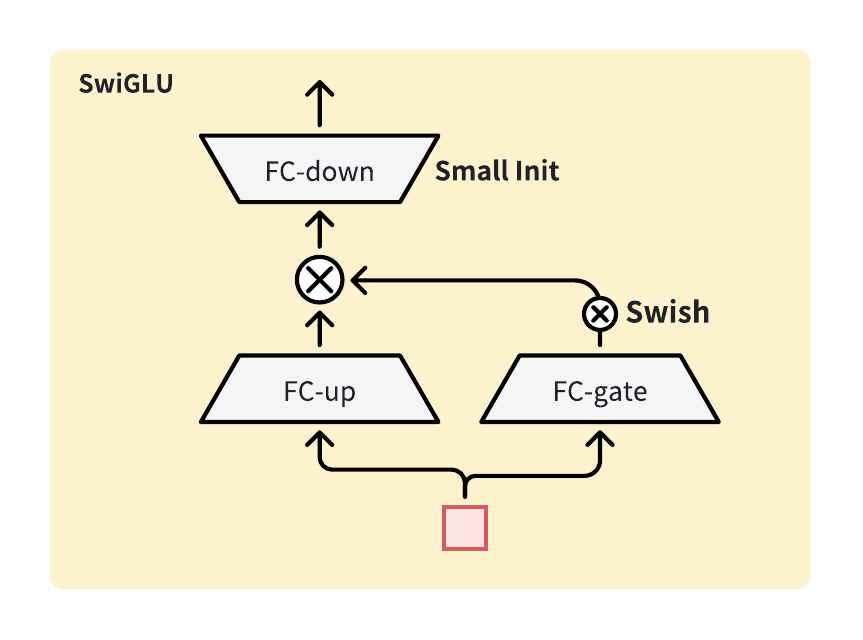}
    \vspace{-0.3cm}
    \captionsetup{skip=-2pt}
    \caption{Small Init in SwiGLU}
    \label{fig:small_init}
\end{figure}

Considering we use Xavier Normal Distribution initialization, which can be formulated as follows:
\begin{equation}
\begin{aligned}
W \sim \mathcal{N}\left(0, \frac{2\cdot scale}{n_{in} + n_{out}}\right)
\end{aligned}
\end{equation}

where $W$ is the kernal of $\mathrm{FCs}$, $n_{in}/ n_{out}$ is the number of input and output units and $scale$ is default set to \textbf{1}. We directly reduced the initialization scale of the final down-projection matrix $\mathrm{FC_{down}}$ in the SwiGLU block to \textbf{0.01}, and keep $\mathrm{FC_{up}/FC_{gate}}$ layers at \textbf{1}. As shown in Figure~\ref{fig:small_init}, this adjustment makes the output $\mathrm{F(x) + x}$ behave more closely to an identity mapping in the early training stages, thereby facilitating the convergence of deep models. Furthermore, the operation $\mathrm{FC_{gate}(x) * Swish(FC_{gate}(x))}$ in the SwiGLU structure tends to amplify the magnitude of activations and gradients in the intermediate hidden layer, which can lead to local instability. Reducing the initialization scale of $\mathrm{FC_{down}}$ also helps constrain the scale of these intermediate representations, enhancing training stability. 

As shown in the table~\ref{tab:small_init_location}, we rigorously evaluated small initialization across different scales (from 1 to 0.01) and different positions (from all layers to reverse ordering). The results demonstrate that applying small initialization specifically to $FC_{down}$ yields the best performance.

\begin{table}[htbp]
  \centering
  \begin{threeparttable}
    \caption{Small Initialization Analysis}
    \label{tab:small_init_location}
    \begin{tabular}{llc}
      \toprule
      \textbf{Version} & \textbf{Stddev Init Value}\tnote{*} & \textbf{$\Delta$AUC} \\
      \midrule
      Base                           &  [1, 1, 1]                      & --            \\
      SmallInit-001                  &  [1, 1, 0.01]                  & +0.03\%   \\
      SmallInit-01                   &  [1, 1, 0.1]                   & +0.02\%           \\
      SmallInit-001-All              &  [0.01, 0.01, 0.01]            & -0.10\%              \\  SmallInit-001-Reverse          &  [0.01, 0.01, 1]            & -0.01\% \\
      \bottomrule
    \end{tabular}
        \begin{tablenotes}[para,flushleft] 
      * denotes $\mathrm{[FC_{up},~FC_{gate},~FC_{down}]}$.
    \end{tablenotes}
  \end{threeparttable}
\end{table}

%% file: software.bib
@article{kaplan2020scaling,
  title={Scaling laws for neural language models},
  author={Kaplan, Jared and McCandlish, Sam and Henighan, Tom and Brown, Tom B and Chess, Benjamin and Child, Rewon and Gray, Scott and Radford, Alec and Wu, Jeffrey and Amodei, Dario},
  journal={arXiv preprint arXiv:2001.08361},
  year={2020}
}

@inproceedings{zhaiactions,
  title={Actions Speak Louder than Words: Trillion-Parameter Sequential Transducers for Generative Recommendations},
  author={Zhai, Jiaqi and Liao, Lucy and Liu, Xing and Wang, Yueming and Li, Rui and Cao, Xuan and Gao, Leon and Gong, Zhaojie and Gu, Fangda and He, Jiayuan and others},
  booktitle={Forty-first International Conference on Machine Learning},
  year={2024}
}

@article{wu2024performance,
  title={Performance Law of Large Language Models},
  author={Wu, Chuhan and Tang, Ruiming},
  journal={arXiv preprint arXiv:2408.09895},
  year={2024}
}

@inproceedings{zhangwukong,
  title={Wukong: Towards a Scaling Law for Large-Scale Recommendation},
  author={Zhang, Buyun and Luo, Liang and Chen, Yuxin and Nie, Jade and Liu, Xi and Li, Shen and Zhao, Yanli and Hao, Yuchen and Yao, Yantao and Wen, Ellie Dingqiao and others},
  booktitle={Forty-first International Conference on Machine Learning},
  year={2024}
}

@article{achiam2023gpt,
  title={Gpt-4 technical report},
  author={Achiam, Josh and Adler, Steven and Agarwal, Sandhini and Ahmad, Lama and Akkaya, Ilge and Aleman, Florencia Leoni and Almeida, Diogo and Altenschmidt, Janko and Altman, Sam and Anadkat, Shyamal and others},
  journal={arXiv preprint arXiv:2303.08774},
  year={2023}
}

@article{dosovitskiy2020image,
  title={An image is worth 16x16 words: Transformers for image recognition at scale},
  author={Dosovitskiy, Alexey},
  journal={arXiv preprint arXiv:2010.11929},
  year={2020}
}

@inproceedings{ramesh2021zero,
  title={Zero-shot text-to-image generation},
  author={Ramesh, Aditya and Pavlov, Mikhail and Goh, Gabriel and Gray, Scott and Voss, Chelsea and Radford, Alec and Chen, Mark and Sutskever, Ilya},
  booktitle={International conference on machine learning},
  pages={8821--8831},
  year={2021},
  organization={Pmlr}
}

@inproceedings{lian2018xdeepfm,
  title={xdeepfm: Combining explicit and implicit feature interactions for recommender systems},
  author={Lian, Jianxun and Zhou, Xiaohuan and Zhang, Fuzheng and Chen, Zhongxia and Xie, Xing and Sun, Guangzhong},
  booktitle={Proceedings of the 24th ACM SIGKDD international conference on knowledge discovery \& data mining},
  pages={1754--1763},
  year={2018}
}

@incollection{wang2017deep,
  title={Deep \& cross network for ad click predictions},
  author={Wang, Ruoxi and Fu, Bin and Fu, Gang and Wang, Mingliang},
  booktitle={Proceedings of the ADKDD'17},
  pages={1--7},
  year={2017}
}

@inproceedings{wang2021dcn,
  title={Dcn v2: Improved deep \& cross network and practical lessons for web-scale learning to rank systems},
  author={Wang, Ruoxi and Shivanna, Rakesh and Cheng, Derek and Jain, Sagar and Lin, Dong and Hong, Lichan and Chi, Ed},
  booktitle={Proceedings of the web conference 2021},
  pages={1785--1797},
  year={2021}
}

@inproceedings{covington2016deep,
  title={Deep neural networks for youtube recommendations},
  author={Covington, Paul and Adams, Jay and Sargin, Emre},
  booktitle={Proceedings of the 10th ACM conference on recommender systems},
  pages={191--198},
  year={2016}
}

@article{zhang2022dhen,
  title={DHEN: A deep and hierarchical ensemble network for large-scale click-through rate prediction},
  author={Zhang, Buyun and Luo, Liang and Liu, Xi and Li, Jay and Chen, Zeliang and Zhang, Weilin and Wei, Xiaohan and Hao, Yuchen and Tsang, Michael and Wang, Wenjun and others},
  journal={arXiv preprint arXiv:2203.11014},
  year={2022}
}

@article{Wang2024remoe,
  author       = {Ziteng Wang and
                  Jianfei Chen and
                  Jun Zhu},
  title        = {ReMoE: Fully Differentiable Mixture-of-Experts with ReLU Routing},
  journal      = {CoRR},
  volume       = {abs/2412.14711},
  year         = {2024},
  url          = {https://doi.org/10.48550/arXiv.2412.14711},
  doi          = {10.48550/ARXIV.2412.14711},
  eprinttype    = {arXiv},
  eprint       = {2412.14711},
  bibsource    = {dblp computer science bibliography, https://dblp.org}
}

@inproceedings{transformer,
 author = {Vaswani, Ashish and Shazeer, Noam and Parmar, Niki and Uszkoreit, Jakob and Jones, Llion and Gomez, Aidan N and Kaiser, \L ukasz and Polosukhin, Illia},
 booktitle = {Advances in Neural Information Processing Systems},
 editor = {I. Guyon and U. Von Luxburg and S. Bengio and H. Wallach and R. Fergus and S. Vishwanathan and R. Garnett},
 pages = {},
 publisher = {Curran Associates, Inc.},
 title = {Attention is All you Need},
 url = {https://proceedings.neurips.cc/paper_files/paper/2017/file/3f5ee243547dee91fbd053c1c4a845aa-Paper.pdf},
 volume = {30},
 year = {2017}
}

@inproceedings{autoint,
author = {Song, Weiping and Shi, Chence and Xiao, Zhiping and Duan, Zhijian and Xu, Yewen and Zhang, Ming and Tang, Jian},
title = {AutoInt: Automatic Feature Interaction Learning via Self-Attentive Neural Networks},
year = {2019},
isbn = {9781450369763},
publisher = {Association for Computing Machinery},
address = {New York, NY, USA},
url = {https://doi.org/10.1145/3357384.3357925},
doi = {10.1145/3357384.3357925},
booktitle = {Proceedings of the 28th ACM International Conference on Information and Knowledge Management},
pages = {1161–1170},
numpages = {10},
location = {Beijing, China},
series = {CIKM '19}
}

@inproceedings{din,
author = {Zhou, Guorui and Zhu, Xiaoqiang and Song, Chenru and Fan, Ying and Zhu, Han and Ma, Xiao and Yan, Yanghui and Jin, Junqi and Li, Han and Gai, Kun},
title = {Deep Interest Network for Click-Through Rate Prediction},
year = {2018},
isbn = {9781450355520},
publisher = {Association for Computing Machinery},
address = {New York, NY, USA},
url = {https://doi.org/10.1145/3219819.3219823},
doi = {10.1145/3219819.3219823},
booktitle = {Proceedings of the 24th ACM SIGKDD International Conference on Knowledge Discovery \& Data Mining},
pages = {1059–1068},
numpages = {10},
location = {London, United Kingdom},
series = {KDD '18}
}

@inproceedings{cheng2016wide,
  title={Wide \& deep learning for recommender systems},
  author={Cheng, Heng-Tze and Koc, Levent and Harmsen, Jeremiah and Shaked, Tal and Chandra, Tushar and Aradhye, Hrishi and Anderson, Glen and Corrado, Greg and Chai, Wei and Ispir, Mustafa and others},
  booktitle={Proceedings of the 1st workshop on deep learning for recommender systems},
  pages={7--10},
  year={2016}
}

@article{guo2017deepfm,
  title={DeepFM: a factorization-machine based neural network for CTR prediction},
  author={Guo, Huifeng and Tang, Ruiming and Ye, Yunming and Li, Zhenguo and He, Xiuqiang},
  journal={arXiv preprint arXiv:1703.04247},
  year={2017}
}

@article{qu2018product,
  title={Product-based neural networks for user response prediction over multi-field categorical data},
  author={Qu, Yanru and Fang, Bohui and Zhang, Weinan and Tang, Ruiming and Niu, Minzhe and Guo, Huifeng and Yu, Yong and He, Xiuqiang},
  journal={ACM Transactions on Information Systems (TOIS)},
  volume={37},
  number={1},
  pages={1--35},
  year={2018},
  publisher={ACM New York, NY, USA}
}

@article{gui2023hiformer,
  title={Hiformer: Heterogeneous Feature Interactions Learning with Transformers for Recommender Systems},
  author={Gui, Huan and Wang, Ruoxi and Yin, Ke and Jin, Long and Kula, Maciej and Xu, Taibai and Hong, Lichan and Chi, Ed H},
  journal={arXiv preprint arXiv:2311.05884},
  year={2023}
}

@inproceedings{pi2020search,
  title={Search-based user interest modeling with lifelong sequential behavior data for click-through rate prediction},
  author={Pi, Qi and Zhou, Guorui and Zhang, Yujing and Wang, Zhe and Ren, Lejian and Fan, Ying and Zhu, Xiaoqiang and Gai, Kun},
  booktitle={Proceedings of the 29th ACM International Conference on Information \& Knowledge Management},
  pages={2685--2692},
  year={2020}
}

@inproceedings{chang2023twin,
  title={TWIN: TWo-stage interest network for lifelong user behavior modeling in CTR prediction at kuaishou},
  author={Chang, Jianxin and Zhang, Chenbin and Fu, Zhiyi and Zang, Xiaoxue and Guan, Lin and Lu, Jing and Hui, Yiqun and Leng, Dewei and Niu, Yanan and Song, Yang and others},
  booktitle={Proceedings of the 29th ACM SIGKDD Conference on Knowledge Discovery and Data Mining},
  pages={3785--3794},
  year={2023}
}

@article{lin2023can,
  title={How can recommender systems benefit from large language models: A survey},
  author={Lin, Jianghao and Dai, Xinyi and Xi, Yunjia and Liu, Weiwen and Chen, Bo and Zhang, Hao and Liu, Yong and Wu, Chuhan and Li, Xiangyang and Zhu, Chenxu and others},
  journal={ACM Transactions on Information Systems},
  year={2023},
  publisher={ACM New York, NY}
}

@misc{longer,
      title={LONGER: Scaling Up Long Sequence Modeling in Industrial Recommenders}, 
      author={Zheng Chai and Qin Ren and Xijun Xiao and Huizhi Yang and Bo Han and Sijun Zhang and Di Chen and Hui Lu and Wenlin Zhao and Lele Yu and Xionghang Xie and Shiru Ren and Xiang Sun and Yaocheng Tan and Peng Xu and Yuchao Zheng and Di Wu},
      year={2025},
      eprint={2505.04421},
      archivePrefix={arXiv},
      primaryClass={cs.IR},
      url={https://arxiv.org/abs/2505.04421}, 
}

@inproceedings{zhu2025rankmixer,
  title={Rankmixer: Scaling up ranking models in industrial recommenders},
  author={Zhu, Jie and Fan, Zhifang and Zhu, Xiaoxie and Jiang, Yuchen and Wang, Hangyu and Han, Xintian and Ding, Haoran and Wang, Xinmin and Zhao, Wenlin and Gong, Zhen and others},
  booktitle={Proceedings of the 34th ACM International Conference on Information and Knowledge Management},
  pages={6309--6316},
  year={2025}
}

@article{tolstikhin2021mlp,
  title={Mlp-mixer: An all-mlp architecture for vision},
  author={Tolstikhin, Ilya O and Houlsby, Neil and Kolesnikov, Alexander and Beyer, Lucas and Zhai, Xiaohua and Unterthiner, Thomas and Yung, Jessica and Steiner, Andreas and Keysers, Daniel and Uszkoreit, Jakob and others},
  journal={Advances in neural information processing systems},
  volume={34},
  pages={24261--24272},
  year={2021}
}

@inproceedings{han2025mtgr,
  title={Mtgr: Industrial-scale generative recommendation framework in meituan},
  author={Han, Ruidong and Yin, Bin and Chen, Shangyu and Jiang, He and Jiang, Fei and Li, Xiang and Ma, Chi and Huang, Mincong and Li, Xiaoguang and Jing, Chunzhen and others},
  booktitle={Proceedings of the 34th ACM International Conference on Information and Knowledge Management},
  pages={5731--5738},
  year={2025}
}

@inproceedings{jiang2022adaptive,
  title={Adaptive domain interest network for multi-domain recommendation},
  author={Jiang, Yuchen and Li, Qi and Zhu, Han and Yu, Jinbei and Li, Jin and Xu, Ziru and Dong, Huihui and Zheng, Bo},
  booktitle={Proceedings of the 31st ACM International Conference on Information \& Knowledge Management},
  pages={3212--3221},
  year={2022}
}

@inproceedings{chen2019behavior,
  title={Behavior sequence transformer for e-commerce recommendation in alibaba},
  author={Chen, Qiwei and Zhao, Huan and Li, Wei and Huang, Pipei and Ou, Wenwu},
  booktitle={Proceedings of the 1st international workshop on deep learning practice for high-dimensional sparse data},
  pages={1--4},
  year={2019}
}

@article{liu2024deepseek,
  title={Deepseek-v3 technical report},
  author={Liu, Aixin and Feng, Bei and Xue, Bing and Wang, Bingxuan and Wu, Bochao and Lu, Chengda and Zhao, Chenggang and Deng, Chengqi and Zhang, Chenyu and Ruan, Chong and others},
  journal={arXiv preprint arXiv:2412.19437},
  year={2024}
}

@article{swietojanski2016learning,
  title={Learning hidden unit contributions for unsupervised acoustic model adaptation},
  author={Swietojanski, Pawel and Li, Jinyu and Renals, Steve},
  journal={IEEE/ACM Transactions on Audio, Speech, and Language Processing},
  volume={24},
  number={8},
  pages={1450--1463},
  year={2016},
  publisher={IEEE}
}

@article{fedus2022switch,
  title={Switch transformers: Scaling to trillion parameter models with simple and efficient sparsity},
  author={Fedus, William and Zoph, Barret and Shazeer, Noam},
  journal={Journal of Machine Learning Research},
  volume={23},
  number={120},
  pages={1--39},
  year={2022}
}

@inproceedings{liu2021swin,
  title={Swin transformer: Hierarchical vision transformer using shifted windows},
  author={Liu, Ze and Lin, Yutong and Cao, Yue and Hu, Han and Wei, Yixuan and Zhang, Zheng and Lin, Stephen and Guo, Baining},
  booktitle={Proceedings of the IEEE/CVF international conference on computer vision},
  pages={10012--10022},
  year={2021}
}

@inproceedings{bachlechner2021rezero,
  title={Rezero is all you need: Fast convergence at large depth},
  author={Bachlechner, Thomas and Majumder, Bodhisattwa Prasad and Mao, Henry and Cottrell, Gary and McAuley, Julian},
  booktitle={Uncertainty in Artificial Intelligence},
  pages={1352--1361},
  year={2021},
  organization={PMLR}
}

@article{yan2025scaling,
  title={From Scaling to Structured Expressivity: Rethinking Transformers for CTR Prediction},
  author={Yan, Bencheng and Lei, Yuejie and Zeng, Zhiyuan and Wang, Di and Lin, Kaiyi and Wang, Pengjie and Xu, Jian and Zheng, Bo},
  journal={arXiv preprint arXiv:2511.12081},
  year={2025}
}

@article{rumelhart1986learning,
  title={Learning representations by back-propagating errors},
  author={Rumelhart, David E and Hinton, Geoffrey E and Williams, Ronald J},
  journal={nature},
  volume={323},
  number={6088},
  pages={533--536},
  year={1986},
  publisher={Nature Publishing Group UK London}
}

@article{zhang2019root,
  title={Root mean square layer normalization},
  author={Zhang, Biao and Sennrich, Rico},
  journal={Advances in neural information processing systems},
  volume={32},
  year={2019}
}

@article{dai2024deepseekmoe,
  title={Deepseekmoe: Towards ultimate expert specialization in mixture-of-experts language models},
  author={Dai, Damai and Deng, Chengqi and Zhao, Chenggang and Xu, RX and Gao, Huazuo and Chen, Deli and Li, Jiashi and Zeng, Wangding and Yu, Xingkai and Wu, Yu and others},
  journal={arXiv preprint arXiv:2401.06066},
  year={2024}
}

@article{chowdhery2023palm,
  title={Palm: Scaling language modeling with pathways},
  author={Chowdhery, Aakanksha and Narang, Sharan and Devlin, Jacob and Bosma, Maarten and Mishra, Gaurav and Roberts, Adam and Barham, Paul and Chung, Hyung Won and Sutton, Charles and Gehrmann, Sebastian and others},
  journal={Journal of Machine Learning Research},
  volume={24},
  number={240},
  pages={1--113},
  year={2023}
}

@article{touvron2023llama,
  title={Llama: Open and efficient foundation language models},
  author={Touvron, Hugo and Lavril, Thibaut and Izacard, Gautier and Martinet, Xavier and Lachaux, Marie-Anne and Lacroix, Timoth{\'e}e and Rozi{\`e}re, Baptiste and Goyal, Naman and Hambro, Eric and Azhar, Faisal and others},
  journal={arXiv preprint arXiv:2302.13971},
  year={2023}
}

@article{zoph2022st,
  title={St-moe: Designing stable and transferable sparse expert models},
  author={Zoph, Barret and Bello, Irwan and Kumar, Sameer and Du, Nan and Huang, Yanping and Dean, Jeff and Shazeer, Noam and Fedus, William},
  journal={arXiv preprint arXiv:2202.08906},
  year={2022}
}
